\pdfoutput=1

\documentclass[11pt,twoside,a4paper,cmspaper,final,collab]{cms-tdr}
\def\svnVersion{342467}\def\cmsCernNoTag{CERN-EP/2016-036}\def\cmsCernDate{\today}\def\cmsMessage{Published in Physics Letters B as \href{http://dx.doi.org/10.1016/j.physletb.2016.05.002}{\doi{10.1016/j.physletb.2016.05.002.}}}

\begin{document}\cmsNoteHeader{SUS-15-002}

\hyphenation{had-ron-i-za-tion}
\hyphenation{cal-or-i-me-ter}
\hyphenation{de-vices}
\RCS$Revision: 337627 $
\RCS$HeadURL: svn+ssh://svn.cern.ch/reps/tdr2/papers/SUS-15-002/trunk/SUS-15-002.tex $
\RCS$Id: SUS-15-002.tex 337627 2016-04-08 10:14:15Z alverson $
\newlength\cmsFigWidth
\ifthenelse{\boolean{cms@external}}{\setlength\cmsFigWidth{0.49\textwidth}}{\setlength\cmsFigWidth{0.9\textwidth}}
\ifthenelse{\boolean{cms@external}}{\providecommand{\cmsLeft}{top\xspace}}{\providecommand{\cmsLeft}{left\xspace}}
\ifthenelse{\boolean{cms@external}}{\providecommand{\cmsRight}{bottom\xspace}}{\providecommand{\cmsRight}{right\xspace}}
\ifthenelse{\boolean{cms@external}}{\providecommand{\cmsAppendix}{}}{\providecommand{\cmsAppendix}{Appendix~}}
\newlength\cmsFigWidthDouble\setlength{\cmsFigWidthDouble}{0.75\textwidth}
\newcommand{\MHT}{\ensuremath{H_{\mathrm T}^{\text{miss}}}\xspace}
\newcommand{\htvecmiss}{\ensuremath{{\vec H}_{\mathrm T}^{\text{miss}}}\xspace}
\newcommand{\njets}{\ensuremath{N_{\text{jet}}}\xspace}
\newcommand{\nbjets}{\ensuremath{N_{{\cPqb}\text{-jet}}}\xspace}
\newcommand{\dphimht}{\ensuremath{\Delta\phi_{\MHT,{\mathrm j}_i}}\xspace}
\newcommand{\rqcd}{\ensuremath{R^{\mathrm{QCD}}}\xspace}
\newcommand{\dphii}{\ensuremath{\Delta \phi}\xspace}
\newcommand{\kht}{\ensuremath{K_{\HT,i}^{\text{data}}}\xspace}
\newcommand{\smht}{\ensuremath{S_{\MHT,j}}\xspace}
\newcommand{\smhtsim}{\ensuremath{S_{\MHT,j}^{\text{sim}}}\xspace}
\newcommand{\snjets}{\ensuremath{S_{\njets,k}^{\text{data}}}\xspace}
\newcommand{\mt}{\ensuremath{m_{\mathrm{T}}}\xspace}
\newcommand{\mlsp}{\ensuremath{m_{{\PSGcz}_1}}\xspace}
\newcommand{\mgluino}{\ensuremath{m_{\PSg}}\xspace}
\newcommand{\zll}{\ensuremath{\cPZ\to \ell^{+}\ell^{-}}\xspace}
\newcommand{\gjets}{{{\cPgg}+jets}\xspace}
\newcommand{\ngdata}{\ensuremath{N_{\gamma}^\text{data}}\xspace}
\newcommand{\nznn}{\ensuremath{N_{\znn}^\text{pred}}\xspace}
\newcommand{\rznn}{\ensuremath{\mathcal{R}_{\znn/\gamma}}\xspace}
\newcommand{\rznnsim}{\ensuremath{\mathcal{R}_{\znn/\gamma}^{\text{sim}}}\xspace}
\newcommand\zjets{{{\cPZ}+jets}\xspace}
\newcommand\wjets{{{\PW}+jets}\xspace}
\newcommand\wpj{\wjets}
\newcommand{\znn}{\ensuremath{\cPZ \to \cPgn \cPagn}\xspace}
\newcommand\zlljets{{\ensuremath{\cPZ (\to \ell^{+} \ell^{-} )}+jets}\xspace}
\newcommand\znnjets{{\ensuremath{\cPZ(\to\cPgn \cPagn )}+jets}\xspace}
\providecommand{\tauh}{\ensuremath{\tau_\mathrm{h}}\xspace}
\providecommand{\mt}{\ensuremath{m_\mathrm{T}}\xspace}
\newcommand{\mtop}{\ensuremath{m_{\mathrm{t}}}\xspace}
\newcommand\imini{\ensuremath{I}\xspace}
\newcommand\rcone{\ensuremath{R}\xspace}
\newcommand{\neles}{\ensuremath{N_{\text{electron}}}\xspace}
\newcommand{\nmuons}{\ensuremath{N_{\text{muon}}}\xspace}
\newcommand{\nisomuons}{\ensuremath{N_{\text{isolated tracks}}^{\text{(muon)}}}\xspace}
\newcommand{\nisoeles}{\ensuremath{N_{\text{isolated tracks}}^{\text{(electron)}}}\xspace}
\newcommand{\nisohads}{\ensuremath{N_{\text{isolated tracks}}^{\text{(hadron)}}}\xspace}

\cmsNoteHeader{SUS-15-002}
\title{Search for supersymmetry in the multijet and missing transverse momentum final state in pp collisions at 13\TeV}

\date{\today}

\abstract{
A search for new physics is performed based on all-hadronic events with
large missing transverse momentum
produced in proton-proton collisions at $\sqrt{s}=13\TeV$.
The data sample,
corresponding to an integrated luminosity of 2.3\fbinv,
was collected with the CMS detector at the CERN LHC in 2015.
The data are examined in search regions
of jet multiplicity,
tagged bottom quark jet multiplicity,
missing transverse momentum,
and the scalar sum of jet transverse momenta.
The observed numbers of events in all search regions
are found to be consistent with the expectations
from standard model processes.
Exclusion limits are presented for simplified supersymmetric
models of gluino pair production.
Depending on the assumed gluino decay mechanism,
and for a massless, weakly interacting, lightest neutralino,
lower limits on the gluino mass from 1440 to 1600\GeV are obtained,
significantly extending previous limits.
}

\hypersetup{%
pdfauthor={CMS Collaboration},%
pdftitle={Search for supersymmetry in the multijet and missing transverse momentum final state in pp collisions at 13 TeV},%
pdfsubject={CMS},%
pdfkeywords={CMS, physics, supersymmetry, multijets}}

\maketitle

\section{Introduction}
\label{sec:introduction}

The standard model (SM) of particle physics successfully
describes a wide range of phenomena.
However, in the SM, the Higgs boson mass is unstable to
higher-order corrections,
suggesting that the SM is incomplete.
Many extensions to the SM have been proposed to
provide a more fundamental theory.
Supersymmetry (SUSY)~\cite{Ramond:1971gb,Golfand:1971iw,Neveu:1971rx,
Volkov:1972jx,Wess:1973kz,Wess:1974tw,Fayet:1974pd,Nilles:1983ge},
one such extension,
postulates that each SM particle is paired with a SUSY
partner from which it differs in spin by one-half unit.
As examples, squarks and gluinos are the SUSY
partners of quarks and gluons, respectively,
while neutralinos \PSGcz (charginos \PSGcpm) arise from a mixture of
the SUSY partners of neutral (charged) Higgs and electroweak gauge bosons.
Radiative corrections involving SUSY particles can compensate the
contributions from SM particles and thereby stabilize the Higgs boson mass.
For this cancellation to be
``natural''~\cite{Barbieri:1987fn,Dimopoulos:1995mi,Barbieri:2009ev,Papucci:2011wy},
the top squark, bottom squark, and gluino
must have masses on the order of a few TeV or less,
possibly allowing them to be produced at the CERN LHC.

Amongst SUSY processes,
gluino pair production,
typically yielding four or more hadronic jets in the final state,
has the largest potential cross section,
making it an apt channel for early SUSY searches in
the recently started LHC Run~2.
Furthermore, in R-parity~\cite{bib-rparity} conserving SUSY models,
as are considered here,
the lightest SUSY particle (LSP) is stable and assumed to be weakly interacting,
leading to potentially large undetected, or ``missing'',
transverse momentum.
Supersymmetry events at the LHC might thus be characterized
by significant missing transverse momentum,
numerous jets,
and --- in the context of natural SUSY ---
jets initiated by top and bottom quarks.

This Letter describes a search for gluino pair production
in the all-hadronic final state.
The data,
corresponding to an integrated luminosity of
2.3\fbinv of proton-proton collisions at a center-of-mass energy
of $\sqrt{s}=13\TeV$,
were collected with the CMS detector in 2015,
the initial year of the LHC Run~2.
Recent searches for gluino pair production at $\sqrt{s}=8\TeV$,
based on data collected in LHC Run~1,
are presented in Refs.~\cite{Aad:2015iea,Khachatryan:2015vra,Khachatryan:2015pwa}.
Because of the large mass scales and their all-hadronic nature,
the targeted SUSY events are expected to exhibit large values of \HT,
where \HT is the scalar sum of the
transverse momenta (\pt) of the jets.
As a measure of missing transverse momentum,
we use the variable \MHT,
which is the magnitude of the vector sum of the jet~\pt.
We present a general search for gluino pair production
leading to final states with large \HT, large \MHT,
and large jet multiplicity.
The data are examined in bins of \njets,
\nbjets, \HT, and \MHT,
where \njets is the number of jets
and \nbjets the number of tagged bottom quark jets ({\cPqb} jets).
The search is performed in exclusive bins of these four observables.

\begin{figure*}[thb]
\centering
\includegraphics[width=0.32\linewidth]{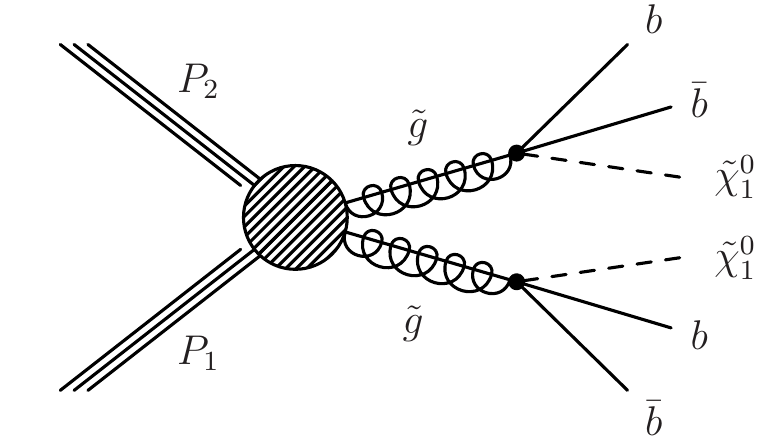}
\includegraphics[width=0.32\linewidth]{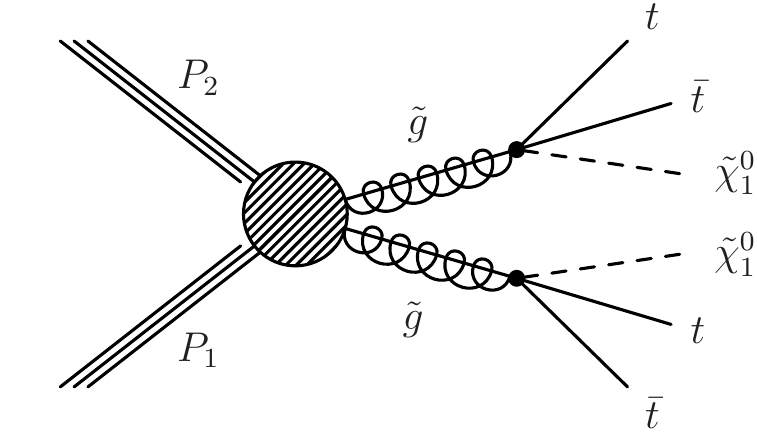}\\
\includegraphics[width=0.32\linewidth]{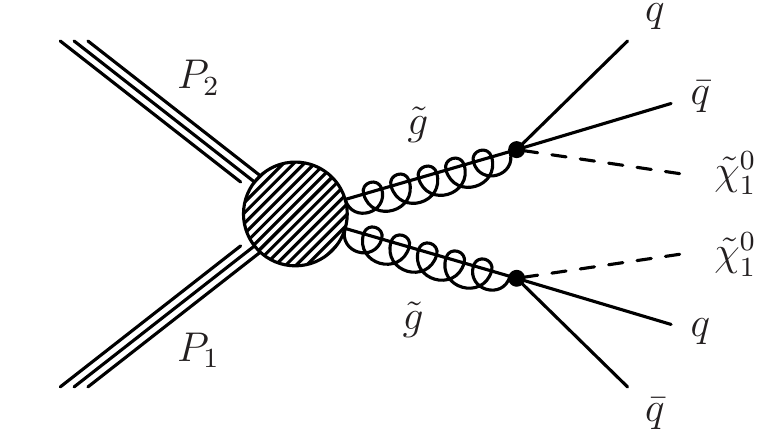}
\includegraphics[width=0.32\linewidth]{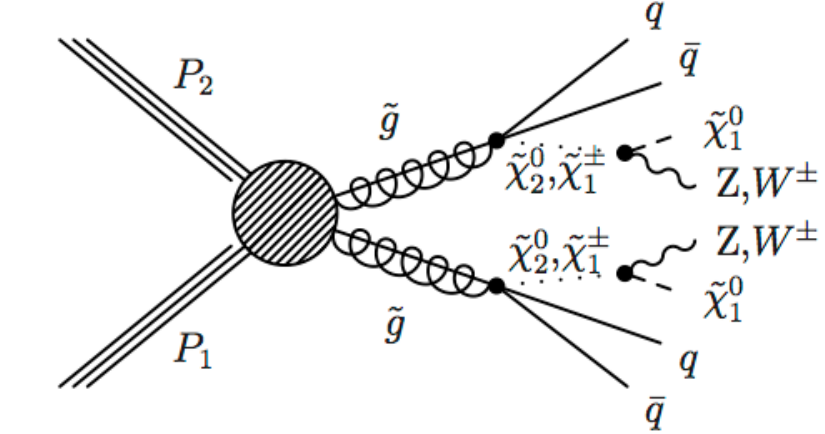}
\caption{
Event diagrams for the new-physics scenarios considered in this study:
the (upper left) T1bbbb, (upper right) T1tttt,
(lower left) T1qqqq, and (lower right) T5qqqqVV
simplified models.
For the T5qqqqVV model,
the quark {\cPq} and antiquark {\cPaq} do not have the same flavor
if the gluino {\PSg} decays as $\PSg\rightarrow\cPq\cPaq\PSGcpm_1$,
with $\PSGcpm_1$ a chargino.
}
\label{fig:event-diagrams}
\end{figure*}

We consider SUSY scenarios in the context of four
simplified models~\cite{bib-sms-1,bib-sms-2,bib-sms-3,bib-sms-4}
of new particle production.
Diagrams for the four models are shown in Fig.~\ref{fig:event-diagrams}.
Simplified models contain the minimal particle content to
represent a topological configuration.
As SUSY production scenarios,
the four simplified models can be interpreted as follows.
In the first scenario,
shown in Fig.~\ref{fig:event-diagrams} (upper left),
gluino pair production is followed by the decay of
each gluino to a bottom quark and an off-shell bottom squark.
The off-shell bottom squark decays to a
bottom quark and the LSP,
where the LSP is assumed  to be the
lightest neutralino \PSGczDo and to escape detection,
leading to significant \MHT.
The second scenario,
shown in Fig.~\ref{fig:event-diagrams} (upper right),
is the same as the first scenario except with top quarks
and off-shell top squarks in place of the bottom quarks and squarks.
The third scenario,
shown in Fig.~\ref{fig:event-diagrams} (lower left),
is the corresponding situation with gluino decay to
a light-flavored quark and off-shell-squark:
up, down, strange, and charm with equal probability,
for each gluino separately.
In the fourth scenario,
shown in Fig.~\ref{fig:event-diagrams} (lower right),
also based on gluino pair production,
each gluino similarly decays to a light-flavored
quark and corresponding off-shell squark.
The off-shell squark decays to a quark and to either the
next-to-lightest neutralino \PSGczDt
or the lightest chargino $\PSGcpm_1$.
The probability for the decay to proceed via the \PSGczDt,
$\PSGcpDo$, or $\PSGc_1^-$,
integrated over the event sample,
is 1/3 for each possibility.
The \PSGczDt ($\PSGcpm_1$) subsequently decays to the \PSGczDo LSP
and to a on- or off-shell {\cPZ} ({$\PW^\pm$}) boson.
We refer to the four simplified models as the T1bbbb, T1tttt,
T1qqqq, and T5qqqqVV scenarios,
respectively~\cite{Chatrchyan:2013sza}.
Thus the first two scenarios explicitly presume
either bottom or top squark production.
The latter two scenarios represent more inclusive situations
and provide complementary
sensitivity to top squark production
for large values of~\njets.
We assume all SUSY particles other than the gluino, the LSP,
and
--- for the T5qqqqVV models ---
the \PSGczDt and $\PSGcpm_1$,
to be too heavy to be directly produced,
and the gluino to be short-lived.

The principal sources of background
arise from the SM production of top quarks,
a {\PW} or {\cPZ} boson in association with jets
(\wjets or \zjets events),
and multiple jets through the strong interaction.
We refer to the latter class of background as
quantum chromodynamics (QCD) multijet events.
The events with top quarks
mostly arise from top quark-antiquark (\ttbar) production,
but also from single top quark processes.
The {\PW} and {\cPZ} bosons
in \wjets and \zjets events can be either on- or off-shell.
For top quark and \wjets events,
significant \MHT can arise if a {\PW} boson decays leptonically,
producing a neutrino and an undetected charged lepton,
while \zjets events can exhibit significant \MHT if
the {\PZ} boson decays to two neutrinos.
For QCD multijet events, significant \MHT can arise if the event
contains a charm or bottom quark that undergoes a semileptonic decay,
but the principal source of \MHT is the mismeasurement of jet~\pt.

This study combines and extends search strategies
developed for the analysis of CMS data collected at $\sqrt{s}=8\TeV$,
specifically the study of Ref.~\cite{Chatrchyan:2013wxa},
which examined data in bins of \nbjets but not \njets
and proved to be sensitive to the T1bbbb scenario,
and the study of Ref.~\cite{Chatrchyan:2014lfa},
which examined data in bins of \njets
but not~\nbjets
and proved to be sensitive to the T1tttt, T1qqqq, and T5qqqqVV scenarios.
Here, the two approaches are combined in a unified framework
to yield a more comprehensive and inclusive study
with improved sensitivity.

\section{Detector, trigger, and event reconstruction}
\label{sec:detector}

The CMS detector is built around a superconducting
solenoid of 6\unit{m} internal diameter,
providing a magnetic field of 3.8\unit{T}.
Within the solenoid volume are a
silicon pixel and strip tracker,
a lead tungstate crystal electromagnetic calorimeter (ECAL),
and a brass and scintillator hadron calorimeter (HCAL).
The ECAL and HCAL, each composed of a barrel and two endcap sections,
extend over a pseudorapidity range $\abs{\eta}<3.0$.
Forward calorimeters on each side of the interaction point
encompass $3.0<\abs{\eta}<5.0$.
The tracking detectors cover $\abs{\eta}<2.5$.
Muons are measured within $\abs{\eta}<2.4$ by gas-ionization detectors embedded
in the steel flux-return yoke outside the solenoid.
The detector is nearly hermetic,
permitting accurate measurements of~\MHT.
A more detailed description of the CMS detector,
together with a definition of the coordinate system and
relevant kinematic variables, is given in Ref.~\cite{Chatrchyan:2008aa}.

Signal event candidates are recorded using trigger conditions based
on thresholds on \HT and missing transverse momentum.
The trigger efficiency,
which exceeds 98\% following application of the event selection
criteria described below,
is measured in data and is accounted for in the analysis.
Separate data samples requiring the presence of either
charged leptons or photons
are used for the determination of backgrounds from SM processes,
as discussed below.

Physics objects are defined using the particle-flow (PF)
algorithm~\cite{cms-pas-pft-09-001,cms-pas-pft-10-001},
which reconstructs and identifies individual particles
through an optimized combination
of information from different detector components.
The PF candidates are classified as photons, charged hadrons, neutral hadrons,
electrons~\cite{Khachatryan:2015hwa},
or muons~\cite{Chatrchyan:2013sba}.
Additional quality criteria are imposed on electron and muon candidates.
For example, more restrictive conditions are placed on the ECAL shower
shape and on the ratio of energies deposited in the HCAL and ECAL for electron
candidates,
and on the matching of track segments between the silicon tracker
and muon detector for muon candidates.
The event primary vertex is taken to be the reconstructed vertex with the
largest sum of charged-track $\pt^2$ values
and is required to lie within 24\unit{cm} (2\unit{cm}) of the
center of the detector
in the direction along (perpendicular to) the beam axis.
Charged tracks from extraneous {\Pp\Pp} interactions within
the same or a nearby bunch crossing (``pileup'')
are removed~\cite{CMS-PAS-JME-14-001}.
The PF objects serve as input for jet reconstruction,
based on the
anti-\kt
algorithm~\cite{Cacciari:2008gp,Cacciari:2011ma}
with a distance parameter of~0.4.
Jet quality criteria as described
in Ref.~\cite{cms-pas-jme-10-003}
are applied to eliminate, for example,
spurious events caused by calorimeter noise.
Contributions to an individual jet's \pt from pileup
interactions are subtracted~\cite{Cacciari:2007fd},
and corrections are applied as a function of jet \pt and~$\eta$
to account for residual effects of
nonuniform detector response~\cite{Chatrchyan:2011ds}.
Jets must have $\pt>30\GeV$.

The identification of {\cPqb} jets is performed by applying
the combined secondary vertex algorithm (CSVv2) at the
medium working point~\cite{CMS-PAS-BTV-15-001}
to reconstructed jets.
The {\cPqb} tagging efficiency
is measured both in a data sample of
multijet events with a reconstructed muon,
and in a data sample of \ttbar events,
with consistent results,
and the probability to misidentify a light-flavor quark or gluon jet
as a {\cPqb} jet in a data sample of inclusive multijet events,
all as a function of jet \pt and~$\eta$.
The signal efficiency for {\cPqb} jets
(misidentification probability for light-flavor quark or gluon jets)
is approximately 55\% (1.6\%)
for jets with $\pt\approx 30\GeV$.
The corresponding misidentification probability for
a charm quark jet is estimated from simulation to be~12\%.

Electrons and muons are required to be isolated in order to
reduce background from events with bottom and charm quarks.
The isolation criterion
is based on the variable~$\imini$,
which is the scalar \pt sum of all PF charged hadrons,
neutral hadrons,
and photons within a cone of
radius $\rcone=\sqrt{\smash[b]{(\Delta\phi)^2+(\Delta\eta)^2}}$
around the lepton direction,
divided by the lepton~\pt,
where $\phi$ is the azimuthal angle.
The sum excludes the lepton under consideration
and is corrected for the contribution of pileup~\cite{CMS-PAS-JME-14-001}.
The cone radius is
$\rcone=0.2$\,(0.05) for lepton $\pt\leq 50\GeV$ ($>$200\GeV),
and $\rcone=10\GeV/\pt$ for $50\leq\pt\leq 200\GeV$.
The reason for the decrease in \rcone with increasing lepton \pt
is to account for the increased collimation of the lepton parent particle's
decay products as the object's Lorentz boost increases.
We require $\imini<0.1$ ($<0.2$) for electrons (muons).

Charged tracks not identified as an isolated electron or muon are
also subjected to an isolation criterion.
To be considered an isolated charged-particle track,
the scalar sum of charged-track \pt values
(excluding the track under consideration)
in a cone of radius $\rcone=0.3$ around the track direction,
divided by the track~\pt,
must be less than~0.2 if the track is identified by the
PF procedure as an electron or muon,
and less than 0.1 otherwise.

\section{Event selection and search regions}
\label{sec:event-selection}

The following requirements define the selection criteria
for signal event candidates:
\begin{itemize}
\item $\njets\geq 4$, where the jets must satisfy $\abs{\eta}<2.4$;
  we require at least four jets because of our focus on gluino
  pair production;
\item $\HT>500\GeV$, where \HT is the scalar \pt sum of jets with $\abs{\eta}<2.4$;
\item $\MHT>200\GeV$, where \MHT is the magnitude of \htvecmiss,
  the negative of the vector \pt sum of jets with $\abs{\eta}<5$;
  the $\eta$ range is extended in this case so that \htvecmiss better represents
  the total missing transverse momentum in an event;
\item no identified, isolated electron or muon candidate with $\pt>10\GeV$;
  electron (muon) candidates are restricted to $\abs{\eta}<2.5$ ($<$2.4);
\item no isolated charged-particle track with $\abs{\eta}<2.4$, $\mt<100\GeV$,
  and $\pt>10\GeV$
  ($\pt>5\GeV$ if the track is identified
  as an electron or muon candidate by the PF algorithm),
  where \mt is the transverse mass~\cite{Arnison:1983rp}
  formed from the \ptvecmiss and isolated-track \pt vector,
  with \ptvecmiss the
  negative of the vector \pt sum of all PF objects;
\item $\dphimht>0.5$ ($>$0.3) for the
  two highest \pt jets j$_1$ and j$_2$
  (the next two highest \pt jets j$_3$ and j$_4$),
  with \dphimht the angle between \htvecmiss
  and the \pt vector of jet j$_i$.
\end{itemize}
The isolated-track requirement eliminates events with
a hadronically decaying $\tau$ lepton,
as well as isolated electrons or muons in cases where the lepton is not identified;
the \mt requirement restricts this veto to tracks consistent
with a {\PW} boson decay
in order to minimize the impact on signal efficiency.
For all-hadronic events, \ptvecmiss and \htvecmiss are similar,
but \htvecmiss is less susceptible to uncertainties in the
modeling of soft energy deposits.
We choose \ptvecmiss for the \mt calculation
for consistency with previous practice.
The \dphimht requirements reduce the background
from QCD multijet processes,
for which \htvecmiss is usually aligned along a jet direction.

\begin{figure}
\centering
    \includegraphics[width=0.49\textwidth]{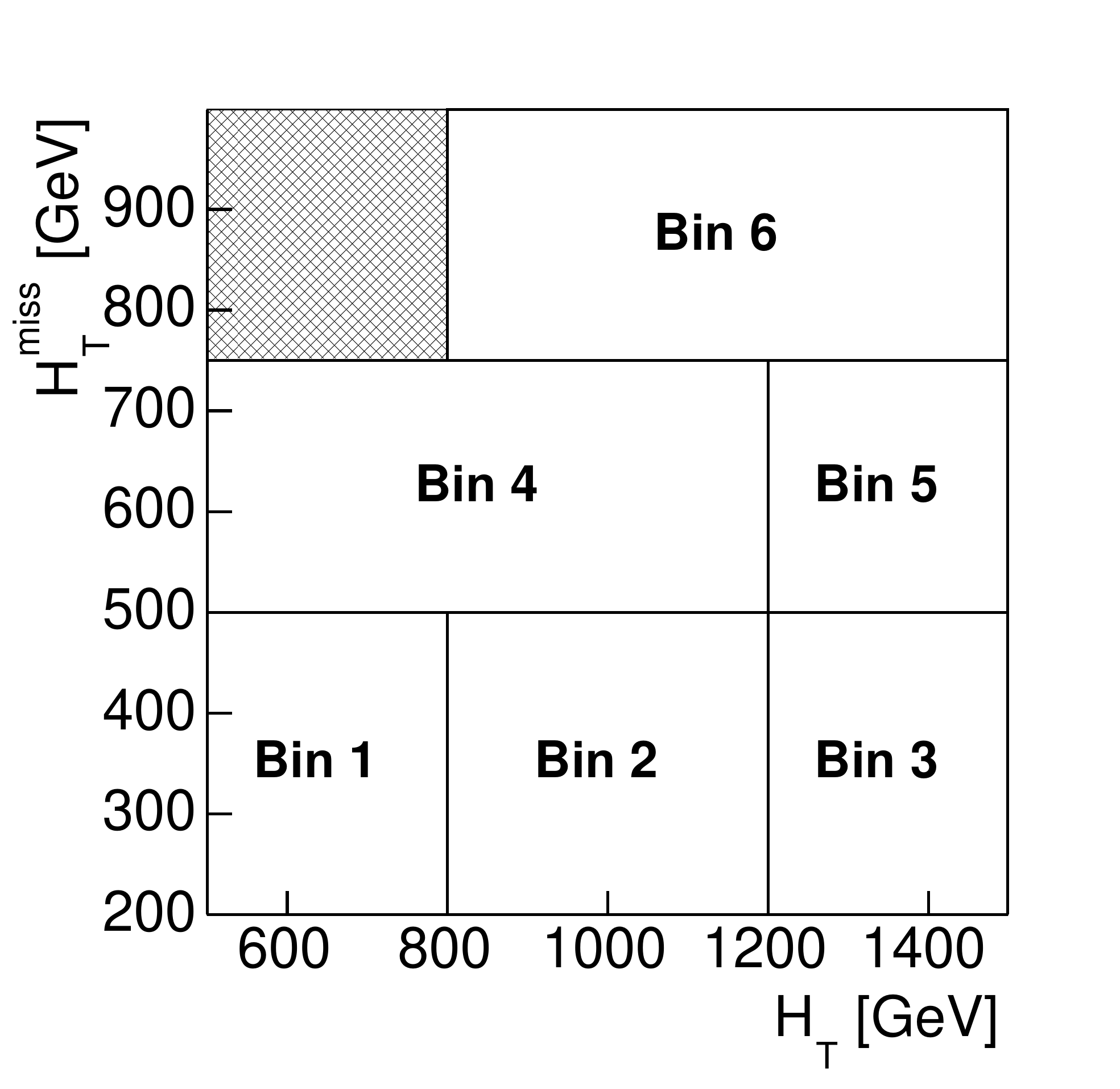}
    \caption{
      Schematic illustration of the search intervals in the \MHT versus \HT plane.
      Each of the six \HT\ and \MHT\ intervals is examined in
      three \njets\ and four \nbjets bins for a total of 72 search regions.
    }
    \label{fig:HT-MHT}
\end{figure}

The search is performed in the following exclusive intervals
of the four search variables:
\begin{itemize}
\item \njets: 4--6, 7--8, $\geq$9;
\item \nbjets: 0, 1, 2, $\geq$3;
\item \HT: 500--800, 800--1200, $\geq$1200\GeV;
\item \MHT: 200--500, 500--750, $\geq$750\GeV.
\end{itemize}
Bins with both $\HT<800\GeV$ and $\MHT>750\GeV$ are discarded because
events with $\MHT\gtrsim\HT$ are very likely to be background.
Additionally, for $500<\MHT<750\GeV$,
an expanded interval $500<\HT<1200\GeV$ is used,
and for $\MHT>750\GeV$ a single interval $\HT>800\GeV$,
because of the low expected number of signal events at large \MHT.
The six search intervals in the \MHT versus \HT plane are
illustrated schematically in Fig.~\ref{fig:HT-MHT}.
The total number of search regions is~72.

A breakdown of the efficiency at different stages of the
selection process for three representative signal models
is given in
\cmsAppendix\ref{sec:sel-eff}.

\section{Event simulation}
\label{sec:mc}

The background is mostly evaluated using data control regions,
as described below (Section~\ref{sec:background}).
Simulated samples of SM events are used to construct
and validate the procedures and to estimate a few of the
smaller background components.
The {\MADGRAPH}5{\textunderscore}a{\MCATNLO} 2.2.2~\cite{Alwall:2014hca}
event generator at leading order
is used to simulate \ttbar, \wjets, \zjets, \gjets,
and QCD multijet events.
This same generator at next-to-leading (NLO) order
is used to describe single top events in the $s$ channel,
events with dibosons
($\PW\PW$, $\cPZ\cPZ$, and $\PW\PH$ production, etc., with $\PH$ a Higgs boson),
and rare processes
($\ttbar\PW$, $\ttbar\cPZ$, and $\PW\PW\cPZ$ production, etc.),
except $\PW\PW$ events in which both {\PW} bosons decay leptonically
are described with the \POWHEG
v1.0~\cite{Nason:2004rx,Frixione:2007vw,Alioli:2010xd,Alioli:2009je,Re:2010bp}
program at NLO.
Single top events in the $t$ and $\cPqt\PW$ channels are
also described with \POWHEG at NLO.
Simulation of the detector response
is based on the \GEANTfour~\cite{Agostinelli:2002hh} package.
The simulated samples are normalized using the most accurate
cross section calculations currently
available~\cite{Alioli:2009je,Re:2010bp,Alwall:2014hca,Melia:2011tj,Beneke:2011mq,
Cacciari:2011hy,Baernreuther:2012ws,Czakon:2012zr,Czakon:2012pz,Czakon:2013goa,
Gavin:2012sy,Gavin:2010az},
generally with NLO or next-to-NLO accuracy.

Signal T1bbbb, T1tttt, T1qqqq, and T5qqqqVV events are generated
for a range of gluino \mgluino and LSP \mlsp mass values,
with $\mlsp<\mgluino$.
For the T5qqqqVV model,
the masses of the intermediate \PSGczDt and $\PSGcpm_1$ states
are taken to be the mean of \mlsp and $\mgluino$.
The signal samples are generated with the
{\MADGRAPH}5{\textunderscore}a{\MCATNLO} program at leading order,
with up to two partons present in addition to the gluino pair.
The decays of the gluino are described with a pure
phase-space matrix element~\cite{Sjostrand:2014zea}.
The signal production cross sections are
computed~\cite{bib-nlo-nll-01,bib-nlo-nll-02,bib-nlo-nll-03,bib-nlo-nll-04,bib-nlo-nll-05}
with NLO plus next-to-leading-logarithm (NLL) accuracy.
To reduce computational requirements,
the detector is modeled with the CMS fast simulation
program~\cite{Orbaker:2010zz,bib-cms-fastsim-02},
which yields consistent results compared with the
{\GEANTfour}-based simulation,
except that we apply a correction of 1\% to account for
differences in the efficiency of the
jet quality requirements~\cite{cms-pas-jme-10-003},
and corrections of 3--10\% to account for differences
in the {\cPqb} jet tagging efficiency.

The NNPDF3.0LO~\cite{Ball:2014uwa}
parton distribution functions (PDF) are used for
the simulated samples generated at leading order,
and the NNPDF3.0NLO~\cite{Ball:2014uwa} PDFs
for the samples generated at NLO.
All simulated samples use the \PYTHIA 8.2~\cite{Sjostrand:2014zea} program
to describe parton showering and hadronization.
To model the effects of pileup,
the simulated events are generated with a nominal
distribution of $\Pp\Pp$ interactions per bunch crossing
and then reweighted to match the
corresponding distribution in data.

\begin{table*}[th]
\topcaption{
Summary of systematic uncertainties that affect the
signal event selection efficiency.
The results are averaged over all search regions.
The variations correspond to different signal
models and choices of the gluino and LSP masses.
}
\centering
\begin{tabular}{lc}
\hline
Item & Relative uncertainty (\%) \\
\hline
Trigger efficiency                              & 0.5--1.1 \\
Pileup reweighting                              & 0.1--0.5 \\
Jet quality requirements                        & 1.0 \\
Renormalization and factorization scales        & 0.1--3.0 \\
Initial-state radiation                         & 0.02--10.0 \\
Jet energy scale                                & 0.5--4.0 \\
Isolated lepton and track vetoes (T1tttt and T5qqqqVV only) & 2.0 \\
\hline
Total                                           & 1.5--11.0 \\
\hline
\end{tabular}
\label{tab:sig-syst}
\end{table*}

We evaluate systematic uncertainties
in the signal model predictions.
Those that are relevant for the selection efficiency
are listed in Table~\ref{tab:sig-syst}.
The uncertainty associated with the renormalization and factorization scales
is determined by varying each scale independently
by factors of 2.0 and 0.5~\cite{Catani:2003zt,Cacciari:2003fi}.
An uncertainty related to the modeling of initial-state radiation (ISR)
is determined by comparing the simulated and measured \pt spectra
of the system recoiling against the ISR jets in \ttbar events,
using the technique described in Ref.~\cite{Chatrchyan:2013xna}.
The two spectra are observed to agree.
The statistical precision of the comparison is used to define
an uncertainty of 15\% (30\%) for $400<\pt<600\GeV$
($\pt>600\GeV$),
while no uncertainty is deemed necessary for $\pt<400\GeV$.
The uncertainties associated with the
renormalization and factorization scales,
and with ISR,
integrated over all search regions,
typically lie below 0.1\%
but can be as large as 1--3\%, and 3--10\%,
respectively,
for $\mlsp\sim\mgluino$
(we use the notation $\mlsp\sim\mgluino$ to
mean $\mlsp + 2m_{\text{X}}\approx\mgluino$,
with $m_{\text{X}}$ the bottom quark mass, the top quark mass,
or the mass of the ``V'' boson, respectively,
for the T1bbbb, T1tttt, and T5qqqqVV models;
for the T1qqqq model,
$\mlsp\sim\mgluino$ means $\mlsp\approx\mgluino$).
The uncertainty associated with the jet energy scale
is evaluated as a function of jet \pt and~$\eta$.
Note that the isolated lepton and track vetoes have a
minimal impact on the T1bbbb and T1qqqq models because events in
these models rarely contain an isolated lepton,
and that the associated uncertainty is negligible ($\lesssim$0.1\%).

We also evaluate systematic uncertainties in the signal predictions
related to the {\cPqb} jet tagging and misidentification efficiencies
and to the statistical uncertainties in the signal event samples.
These sources of uncertainty do not affect the signal
efficiency but can potentially alter the signal distribution shapes.
Similarly,
the sources of systematic uncertainty associated with the trigger efficiency,
pileup reweighting,
renormalization and factorization scales, ISR,
and jet energy scale
can affect the shapes of the signal distributions.
These potential changes in shape,
i.e., migration of events between search regions,
are accounted for in the limit-setting procedure
described in Section~\ref{sec:results}.

The systematic uncertainty in the determination
of the integrated luminosity is 4.6\%.

\section{Background evaluation}
\label{sec:background}

In this section,
we describe the evaluation of the background from SM processes.
This evaluation relies on data control regions (CRs)
selected using similar criteria to the search regions.
Signal events may contribute to the CRs.
The impact of this ``signal contamination'' on the final results
is evaluated in the context of each individual SUSY model,
as described in Section~\ref{sec:results}.
However, the level of signal contamination
is negligible for all CRs except those used to evaluate the
top quark and \wjets background (Section~\ref{sec:ttbar}),
and is nonnegligible only for the T1tttt and T5qqqqVV models.
The level of signal contamination for these nonnegligible
cases is discussed in Sections~\ref{sec:ttbar-ll}
and~\ref{sec:ttbar-hadtau}.

\subsection{Background from top quark and \texorpdfstring{\wjets}{W+jets} events}
\label{sec:ttbar}

Background from SM \ttbar, single top quark, and \wjets events
arises when a {\PW} boson decays leptonically,
yielding a neutrino (thus, genuine \MHT) and a non-vetoed charged lepton.
The non-vetoed lepton can be an electron or muon
(including from $\tau$ lepton decay)
that does not satisfy the identification requirements
of Section~\ref{sec:event-selection}
(so-called ``lost leptons''),
or it can be a hadronically decaying $\tau$ lepton.

\subsubsection{Lost-lepton background}
\label{sec:ttbar-ll}

Lost-lepton background can arise
if an electron or muon lies outside the analysis acceptance,
is not isolated, or is not reconstructed.
The lost-lepton background is evaluated following
the procedures established in
Refs.~\cite{Chatrchyan:2014lfa,Collaboration:2011ida,Chatrchyan:2012lia}.
Briefly,
single-lepton CRs are selected
by inverting the electron and muon vetoes.
Each CR event is entered into one of the 72 search regions
with a weight that represents
the probability for a lost-lepton event to appear with
the corresponding values of \HT, \MHT, \njets, and \nbjets.

The CRs are selected by requiring events to satisfy
the criteria of Section~\ref{sec:event-selection}
except exactly one isolated electron or muon must be present
and the isolated-track veto is not applied.
The transverse mass formed from the \ptvecmiss and
lepton \pt vector is required to satisfy $\mt<100\GeV$:
this requirement is effective at identifying SM events,
which primarily arise from leptonic {\PW} boson decay,
while reducing signal contamination.
After applying this requirement,
the fraction of CR events due to T1tttt (T5qqqqVV)
signal contamination is generally negligible,
viz., $\lesssim 0.1\%$,
but it can be as high as around 30--40\% (5--20\%)
for the largest values of  \njets, \nbjets, \HT, and/or \MHT,
depending on \mgluino and~\mlsp.
The weights,
accounting for the probability for a lepton to be ``lost'',
are determined from the \ttbar, \wjets, single top quark,
and rare process simulations
through evaluation of the efficiency of the acceptance,
reconstruction, and isolation requirements
as a function of \HT, \MHT, \njets,
lepton \pt, and other kinematic variables.
Since the efficiencies are parametrized
in terms of kinematic and topological quantities,
the method is insensitive to the specific mix of processes,
i.e., it does not require
the relative fractions of \ttbar, single top, and \wjets events
in the CRs to be the same as in the search regions
(nonetheless, these fractions agree to within less than 1\% in simulation).
A correction derived from data is
applied to the weights to account for the trigger efficiency,
while corrections from simulation account for
contamination due to nonprompt electrons,
contamination due to dilepton events in which one of the leptons is lost,
and the selection efficiency of the {\mt} requirement.
Corresponding efficiencies are evaluated for dileptonic events
in which both leptons are lost.
This latter source of background
is predicted to account for $<$2\% of the total lost-lepton background.
Finally, a correction is applied to account for the selection
efficiency of the isolated-track veto.

\begin{figure*}[htp]
  \centering
  \includegraphics[width=\cmsFigWidthDouble]{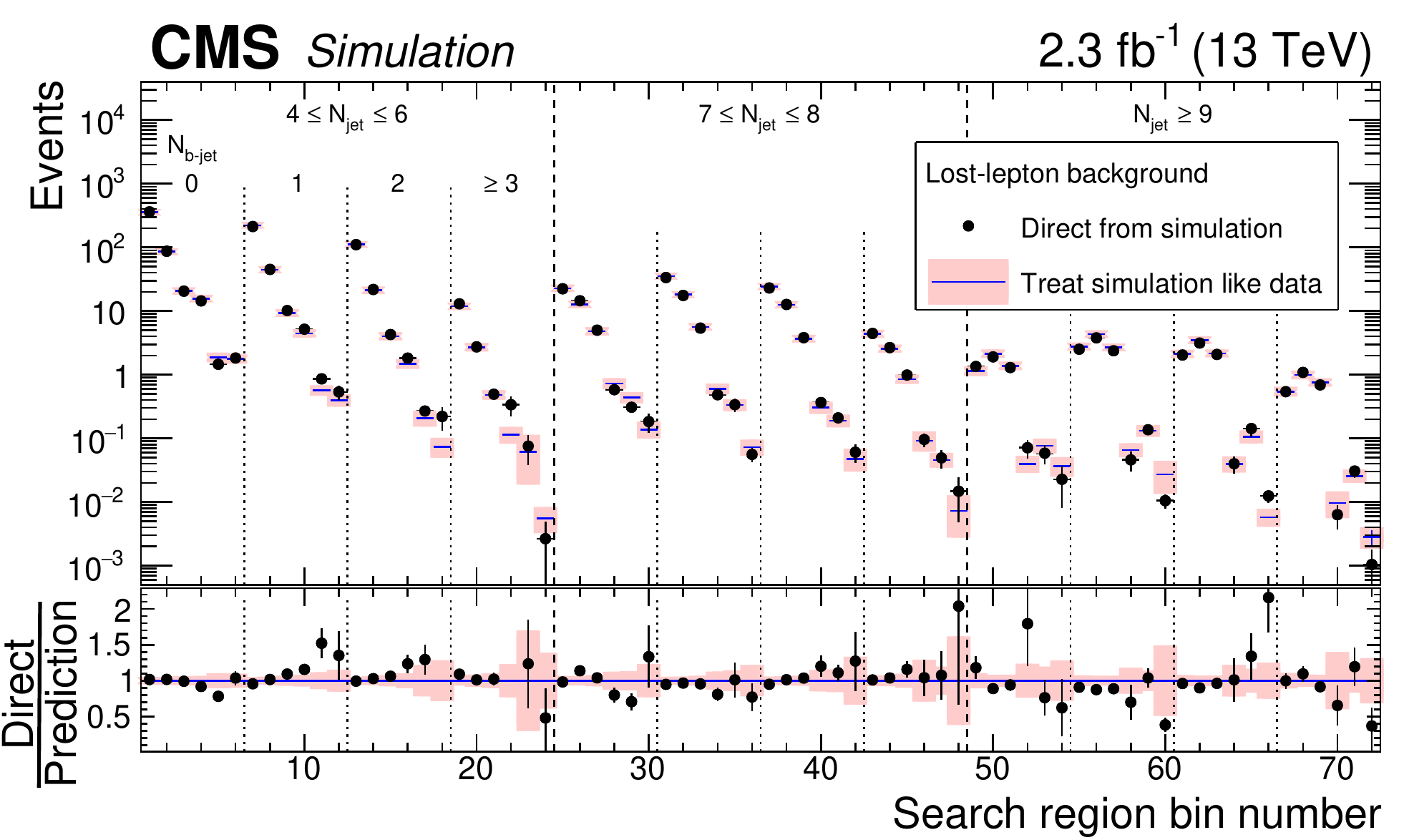}
  \includegraphics[width=\cmsFigWidthDouble]{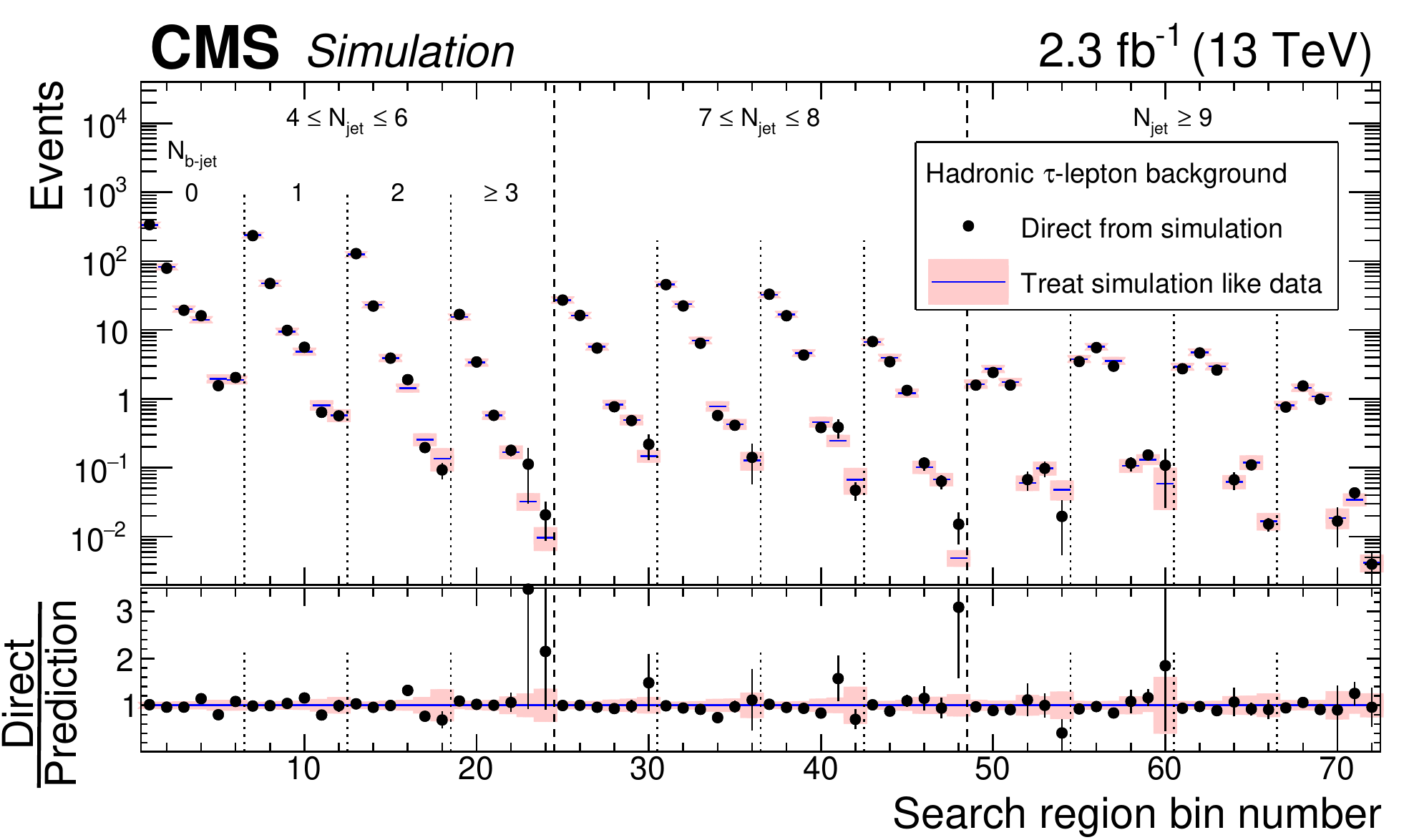}
  \caption{(upper plot) The lost-lepton background in the 72 search regions
    of the analysis as determined directly from \ttbar,
    single top quark, \wjets, diboson, and rare-event simulation
    (points, with statistical uncertainties) and as predicted by applying the
    lost-lepton background determination procedure to simulated
    electron and muon control samples
    (histograms, with statistical uncertainties).
    The lower panel shows the same results following division
    by the predicted value,
    where bins without markers have ratio values outside the scale of the plot.
    (lower plot) The corresponding simulated results for the background from
    hadronically decaying $\tau$ leptons.
    For both plots, the six results within each region delineated by
    dashed lines
    correspond sequentially to the six regions of \HT and \MHT
    indicated in Fig.~\ref{fig:HT-MHT}.
  }
  \label{fig:lost-lepton-closure}
\end{figure*}

The weighted distributions of the search variables,
summed over the events in the CRs,
define the lost-lepton background prediction.
The procedure
is performed separately for single-electron
and single-muon events.
The two independent predictions yield consistent results and are
averaged to obtain the final lost-lepton background prediction.
The method is validated with a closure test,
namely by determining the ability of the method,
applied to simulated samples,
to predict correctly the true number of background events.
The results of the closure test are shown
in the upper plot of Fig.~\ref{fig:lost-lepton-closure}.
As a check,
we repeated the closure test after varying the
fractions of \ttbar, single top, and \wjets events,
with no discernible change in the outcome.

The dominant uncertainties in the lost-lepton background
prediction are statistical,
due to the limited number of CR events in the most sensitive search regions.
As a systematic uncertainty,
we take the larger of the observed nonclosure
in Fig.~\ref{fig:lost-lepton-closure}~(upper plot)
or the statistical uncertainty in the nonclosure,
for each search region,
where ``nonclosure'' refers to the difference between
the solid points and histogram.
Additional systematic uncertainties are assigned
based on a comparison
between data and simulation of the lepton reconstruction,
lepton isolation,
and isolated track veto efficiencies.
Within the statistical precision,
there are no such differences observed,
and the statistical uncertainty in the respective
comparison is assigned as a systematic uncertainty.
Uncertainties in the acceptance associated with the PDFs,
including those related to the renormalization and factorization scales,
are evaluated by varying the PDF sets used to produce the simulated samples.
These uncertainties are defined by the maximum deviations observed
from 100 variations of the
NNPDF3.0LO PDFs for \ttbar and \wjets events.
The uncertainty in the jet energy correction is
propagated to \ptvecmiss,
and the resulting change in the {\mt} selection efficiency
is used to define a systematic uncertainty.
Small systematic uncertainties related to the purity of the
electron and muon CRs
and to the statistical uncertainties
in the simulated efficiencies are also evaluated.

\subsubsection{Hadronically decaying \texorpdfstring{$\tau$}{tau} lepton background}
\label{sec:ttbar-hadtau}

To evaluate the background due to {\PW} bosons
that decay to a neutrino and a hadronically decaying $\tau$ lepton ($\tauh$),
we employ a template
method~\cite{Chatrchyan:2014lfa,Collaboration:2011ida,Chatrchyan:2012lia}.
The \tauh background is determined from a single-muon CR,
composed almost entirely of \ttbar, single top quark, and \wjets events,
selected using a trigger that requires $\HT>350\GeV$ and
at least one muon candidate with $\pt>15\GeV$.
The CR events are required to contain exactly
one identified muon with $\pt>20\GeV$ and $\abs{\eta}<2.1$.
Since $\mu$+jets and $\tauh$+jets production arise from
the same underlying process,
the hadronic component of the events is expected to be the same
aside from the response of the detector to a $\mu$ or $\tauh$.
The muon \pt in the single-muon CR is smeared according to
the response functions (``templates'') derived from \ttbar and \wjets simulation.
The templates express the expected visible-\pt distribution of a \tauh candidate
as a function of the true $\tau$-lepton \pt value,
taken to be the measured muon~\pt.
The fraction of T1tttt (T5qqqqVV) events in the CR due to
signal contamination is generally $\lesssim0.1\%$,
but can be as large as around 15--25\% (4--8\%)
for the largest values of \njets, \nbjets, \HT, and/or \MHT,
depending on \mgluino and~\mlsp.

Following the smearing,
the values of \HT, \MHT, \njets, and \nbjets are calculated for the CR event,
and the selection criteria of Section~\ref{sec:event-selection} are applied.
The misidentification probability for a \tauh jet to be erroneously
identified as a {\cPqb} jet is taken into account.
Corrections are applied to account for the trigger efficiency,
the acceptance and efficiency of the $\mu$ selection,
and the ratio of branching fractions
${BF}(\PW\to\tauh\nu)/{BF}(\PW\to\mu\nu) = 0.65$~\cite{PDG2014}.
The resulting event yield provides the \tauh background estimate.
The method is validated with a closure test,
whose results are shown in the lower plot of Fig.~\ref{fig:lost-lepton-closure}.
Systematic uncertainties are assigned based on the level of closure,
as described for the lost-lepton background.
Other systematic uncertainties
are associated with the muon acceptance,
the response functions,
and the misidentification rate of \tauh jets as {\cPqb}~jets.
The dominant uncertainty,
as for the lost-lepton background,
arises from the limited number of events in the~CR.

\subsection{Background from \texorpdfstring{\znn}{Z -> nu nu} events}
\label{sec:znn}

A straightforward method to evaluate the background from \zjets events
with \znn consists of selecting \zjets events with {\zll}
($\ell=\Pe$, $\Pgm$),
removing the $\ell^+$ and $\ell^-$ to emulate the \znn process,
and applying the event selection criteria of
Section~\ref{sec:event-selection}.
The resulting efficiency-corrected event yields can be
directly translated into a prediction for the \znn background
through multiplication by the
known ratio of branching fractions~\cite{PDG2014}.
A limitation of this procedure is the small {\zll} branching fraction
in relation to that for~\znn.

An alternative approach is to exploit the similarity between \cPZ\ boson
radiation and the more copious radiation of photons
by selecting \gjets events,
removing the photon from the event,
and applying the selection criteria of Section~\ref{sec:event-selection}.
The \gjets process differs from the \zjets process because of threshold effects
associated with the {\cPZ} boson mass
and because of the different couplings of {\cPZ} bosons and
photons to up- and down-type quarks.
These differences are generally well understood and
described adequately with simulation.

Our evaluation of the \znn background utilizes both approaches.
A \gjets CR is selected using a trigger
that requires $\HT>500\GeV$ and photon $\pt>90\GeV$.
A \zjets CR with \zll\ is selected using a trigger that requires
$\HT>350\GeV$ and at least one electron or muon with $\pt>15\GeV$.
Fits as described in Refs.~\cite{Chatrchyan:2014lfa}
and~\cite{Chatrchyan:2013wxa}
are used to extract the prompt-photon and {\cPZ} boson yields,
respectively.
Because of current limitations in the simulations for the theoretical modeling
of \gjets versus \zjets production with heavy flavor jets,
we restrict the use of \gjets events to the 18 search
regions with $\nbjets=0$.
The \zll sample,
integrated over \HT and \MHT
because of the limited statistical precision,
is used to extrapolate the $\nbjets=0$ results
to the $\nbjets>0$ search regions.

The \gjets analysis is similar to that presented
in Ref.~\cite{Chatrchyan:2014lfa}.
We predict the number \nznn of \znnjets events
contributing to each $\nbjets=0$ search region
from the number \ngdata of events in the corresponding
\njets, \HT, and \MHT bin of the \gjets CR:
\begin{equation}
  \left.\nznn\right|_{\nbjets=0}
     = \rho {\rznnsim} \beta_\gamma^{\text{data}} \ngdata,
\label{eq:gjet1}
\end{equation}
where $\beta_\gamma^{\text{data}}$ is the purity of the CR,
determined from the fit~\cite{Chatrchyan:2014lfa} to data,
and \rznnsim the ratio from simulation (``sim'')
of the numbers of \znnjets events to \gjets events,
with the \gjets term obtained from a leading-order
{\MADGRAPH}5{\textunderscore}a{\MCATNLO} calculation.
Corrections are applied to account for efficiency differences between
the data and simulation and for an angular cutoff in the simulation
that controls the singularity associated with soft collinear
radiative corrections.
The factor $\rho$~\cite{Chatrchyan:2014lfa} in Eq.~(\ref{eq:gjet1}), defined as
\begin{equation}
\rho = \frac{\mathcal{R}_{\zll/\gamma}^\text{data}}{\mathcal{R}_{\zll/\gamma}^\text{sim}} = \frac{N^\text{data}_{\zll}}{N^\text{data}_\gamma} \frac{N^\text{sim}_{\gamma}}{N^\text{sim}_{\zll}},
\label{eq:doubleratio}
\end{equation}
uses the \zll CR to account for potential differences in the \rznn
factor between simulation and data,
such as those expected due to missing higher-order terms
in the \gjets calculation,
and is found to have a value of 0.92 (taken to be constant),
with uncertainties,
deduced from linear fits to projections onto each dimension,
that vary with \njets, \HT, and \MHT between 8 and 60\%.

For search regions with $\nbjets>0$,
the \znn background estimate is
\begin{align}
  \left(\nznn\right)_{j,b,k} &= \left(\nznn\right)_{j,0,k}\mathcal{F}_{j,b};  \label{eq:Zbjet} \\
  \mathcal{F}_{j,b} &=
  \left[\left(N^\text{data}_{\zll} \beta_{\ell\ell}^{\text{data}}\right)_{0,b} /
  \left(N^\text{data}_{\zll} \beta_{\ell\ell}^{\text{data}}\right)_{0,0}\right]
  \mathcal{J}_{j,b}; \label{eq:Fbjet} \\
  \mathcal{J}_{j,b} &= N^\text{model}_{j,b}/N^\text{model}_{0,b}, \label{eq:Jbjet}
\end{align}
where $j$, $b$, and $k$ are bin indices (numbered from zero)
for the \njets, \nbjets, and kinematic
(i.e., \HT and \MHT) variables, respectively.
For example, $j=0$ ($b=3$) corresponds to \njets=4--6 ($\nbjets\ge3$),
while $k=0$ denotes ``Bin~1'' of Fig.~\ref{fig:HT-MHT}.
The first term on the right-hand side of Eq.~(\ref{eq:Zbjet}) is
obtained from Eq.~(\ref{eq:gjet1}).
The \nbjets extrapolation factor $\mathcal{F}$ [Eq.~(\ref{eq:Fbjet})]
is obtained from the fitted \zll yields,
with data-derived corrections $\beta_{\ell\ell}^{\text{data}}$
to account for the {\nbjets}-dependent purity.
Other efficiencies cancel in the ratio.
The dependence of the \nbjets shape of $\mathcal{F}$ on \njets is described
with the factor $\mathcal{J}$ [Eq.~(\ref{eq:Jbjet})],
which is determined using a model estimate $N^\text{model}_{j,b}$
because of the limited statistical precision of the \zll data.
The model uses the results of the \zll simulation
for the central value of $\mathcal{J}$.
Based on simulation studies,
we determine corresponding upper and lower bounds
to define a systematic uncertainty.
As a lower bound on $\mathcal{J}$,
we set $N^\text{model}_{j,b}=N^\text{model}_{0,b}$,
i.e., $\mathcal{J}_{j,b}=1$ in Eq.~(\ref{eq:Fbjet}).
In this limit $\mathcal{F}$ is independent of \njets,
corresponding to a factorization of the mechanisms
to produce bottom quark jets and additional jets.
As an upper bound,
we take
$N^\text{model}_{j,b}=\sum_{\njets\in j, \nbjets\in b}{\mathcal{B}(\nbjets|\njets;p)}$,
where $\mathcal{B}$ is a binomial distribution,
with $p$ the probability for a jet to be tagged as a {\cPqb} jet.
In both simulation and data we find $p$ to be independent of \njets.
This binomial behavior would be expected should all tagged b jets be erroneous,
i.e., not initiated by b quarks,
or should the production of quarks in the hadron shower not depend on
flavor except via a scale factor that is absorbed into the empirical factor~$p$.
With respect to a systematic uncertainty,
the factorization and binomial extrapolations represent opposite extremes.
The binomial assumption is validated in simulation;
the result $p=0.062\pm0.007$ is obtained from a fit to the data,
of which $\simeq$0.02 is attributable to light-parton or charm quark jets
erroneously identified as {\cPqb} jets.
The resulting systematic uncertainties in $\mathcal{J}$ range
from a few percent to $\approx$60\%, depending on \njets and \nbjets.

\begin{figure*}[ht]
\centering
\includegraphics[width=\cmsFigWidthDouble]{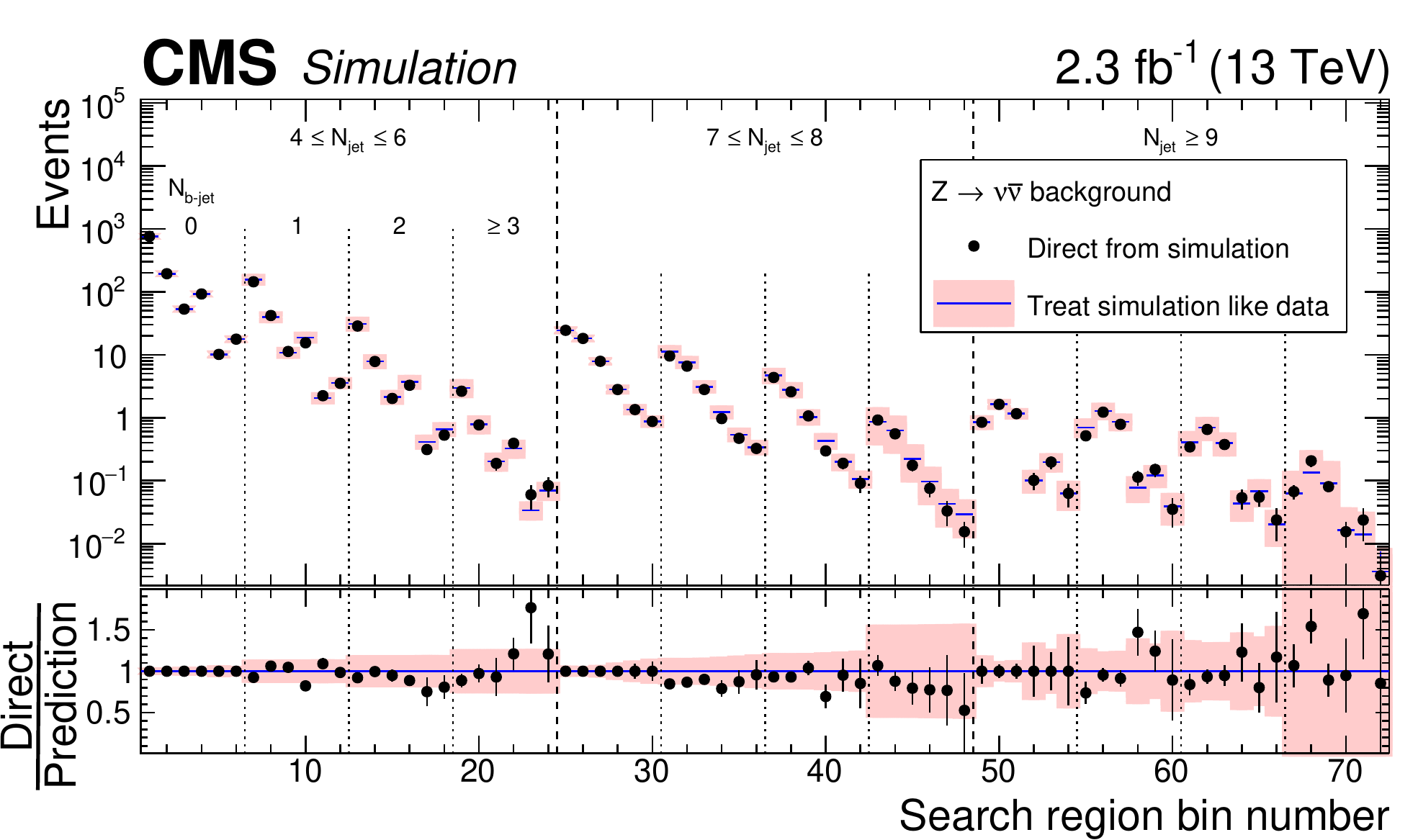}
\caption{
  The \znn background in the 72 search regions of the analysis
  as determined directly from \znnjets and
  $\ttbar\cPZ$ simulation (points),
  and as predicted by applying the \znn background determination
  procedure to statistically independent
  \zlljets simulated event samples (histogram).
  For bins corresponding to $\nbjets=0$,
  the agreement is exact by construction.
  The lower panel shows the ratio between the true and predicted yields.
  For both the upper and lower panels,
  the shaded regions indicate the quadrature sum of the
  systematic uncertainty associated with the
  dependence of $\mathcal{F}$ on the kinematic parameters (\HT and \MHT)
  and the statistical uncertainty of the simulated sample.
  The labeling of the search regions is the same as in Fig.~\ref{fig:lost-lepton-closure}.
}
\label{fig:bjetSF72}
\end{figure*}

A closure test of the method
is presented in Fig.~\ref{fig:bjetSF72}.
The shaded bands represent the systematic uncertainty
(10--20\%, depending on \nbjets)
arising from our treatment of $\mathcal{F}$ as
independent of the kinematic parameters,
combined with the statistical
uncertainty of the \zlljets simulation.

Rare processes such as $\ttbar\cPZ$
and V(V)\cPZ\ ($\mathrm{V}=\PW$ or \cPZ) production
can contribute to the background.
We add the expectations for these processes,
obtained from simulation,
to the background predicted from the procedure described above.
Note that processes with a {\cPZ} boson
and a $\cPZ\to\gamma$ counterpart
are already accounted for in $N^\text{data}_\gamma$
and largely cancel in the {\rznn} ratio.
For search regions with $\nbjets\ge2$,
the contribution of $\ttbar\cPZ$ events
is found to be comparable to that from \zjets events,
with an uncertainty of $\approx$50\%,
consistent with the rate and uncertainty for
$\ttbar\cPZ$ events found in Ref.~\cite{cms_ttZ}.

Besides the uncertainty related to the \nbjets extrapolation,
discussed above,
systematic uncertainties associated with
the statistical precision of the simulation,
the photon reconstruction efficiency,
the photon and dilepton purities,
and the $\rho\rznnsim$ term are evaluated.
Of these,
the $\rho\rznnsim$ term (10--60\%) dominates the overall uncertainty
except in the highest (\njets, \nbjets) search regions where the
overall uncertainty is dominated by the
statistical precision of the simulation (70--110\%)
and by the uncertainty in the \zll purity (40\%).
The underlying source of the leading systematic uncertainties is the
limited number of events in the CR.

\subsection{Background from QCD multijet events}
\label{sec:qcd}

To evaluate the background associated with QCD multijet production,
we select a QCD dominated CR by inverting the \dphimht requirements,
i.e., by requiring at least one of the four highest \pt jets in an event to fail
the respective \dphimht selection criterion
listed in Section~\ref{sec:event-selection}.
The resulting sample is called the ``low-\dphii'' CR.
The QCD background in each search region is given by the product
of the observed event yield in the corresponding region
of the low-\dphii CR
multiplied by a factor \rqcd expressing the ratio of the
expected QCD multijet background in the
respective signal and low-\dphii regions,
taking into account the contributions from non-QCD SM processes.
The non-QCD SM contributions to the low-\dphii CR,
which correspond to around 14\% of the events in this CR,
are evaluated
using the techniques described above for the top quark,
\wjets, and \zjets backgrounds,
except with the inverted \dphimht requirements.
The \rqcd terms are determined primarily from data,
as described below.
The procedure is analogous to that used
in Refs.~\cite{Chatrchyan:2013wxa,Chatrchyan:2012rg}
to evaluate the QCD multijet background.

\begin{figure*}[!htb]
\centering
\includegraphics[width=\cmsFigWidthDouble]{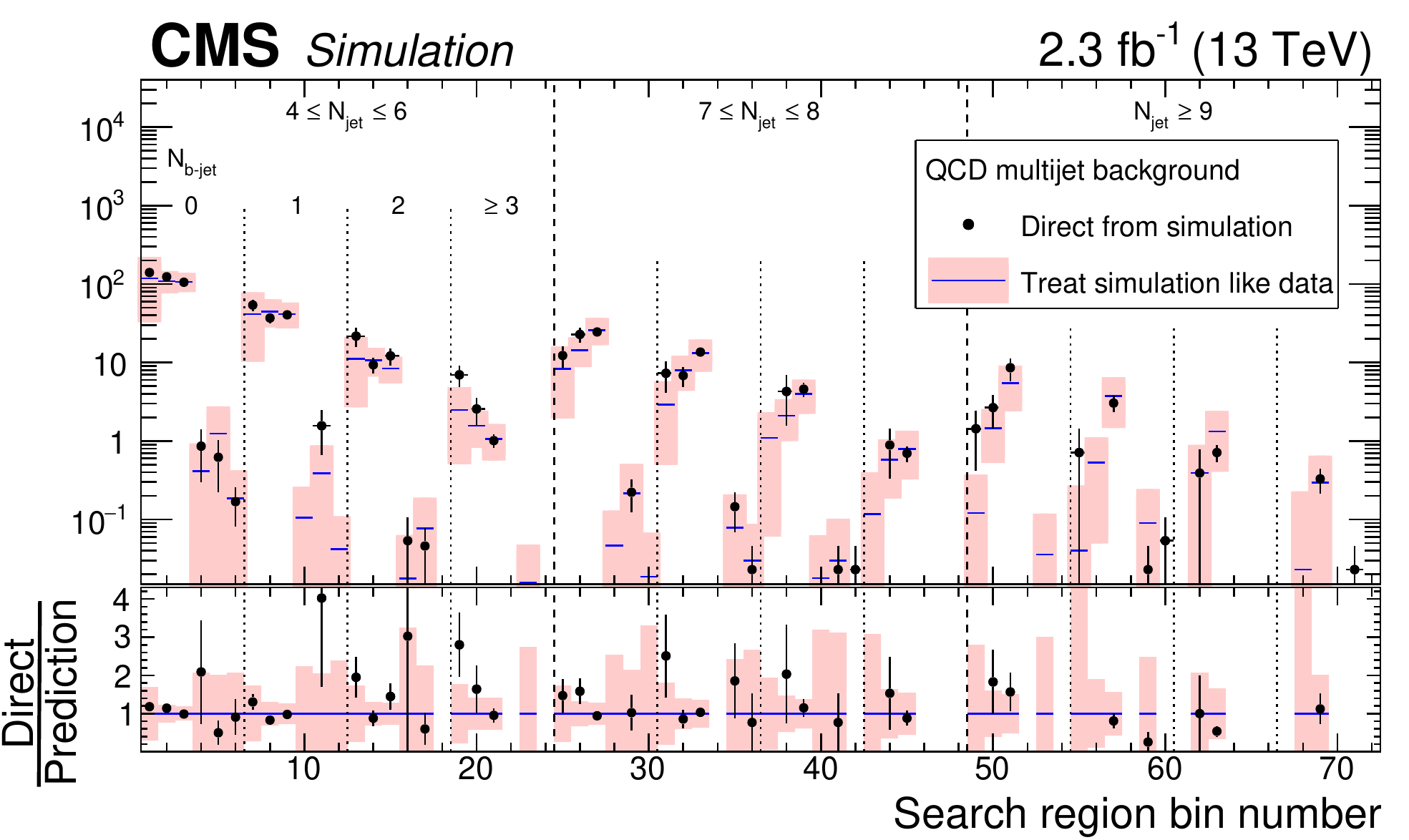}
\caption{
The QCD multijet background in the 72 search regions of the analysis
as determined directly from QCD multijet simulation
(points, with statistical uncertainties)
and as predicted by applying the QCD multijet background determination
procedure to simulated event samples
(histograms, with statistical and systematic uncertainties added in quadrature).
The lower panel shows the same results following division
by the predicted value.
The labeling of the search regions is the same as in Fig.~\ref{fig:lost-lepton-closure}.
Bins without markers have no events in the control regions.
No result is given in the lower panel if the value of
the prediction is zero.
}
\label{fig:qcd-mdp-closure}
\end{figure*}

The \rqcd factor increases with \njets but is found empirically to have a
negligible dependence on \nbjets for a given \njets value.
We therefore divide the $4\leq\njets\leq6$ search region
into three exclusive bins: $\njets=4$, 5, and~6.
Once this is done,
there is no dependence of \rqcd on \nbjets.
Similarly, we divide the $200\leq\MHT\leq 500\GeV$ search region
into two bins:
$200<\MHT<300\GeV$ and $300<\MHT<500\GeV$;
the first of these two bins
is enhanced in QCD background events,
both in the low-\dphii and signal samples.
The \HT, \MHT, and \njets dependence of \rqcd is modeled as:
\begin{equation}
  \label{eqn:rqcd}
  \rqcd_{i,j,k} = \kht \smhtsim  \snjets ,
\end{equation}
where $i$, $j$, and $k$ are bin indices.
The \kht term is the ratio of the expected number of
QCD multijet events in the search region to that in the
low-\dphii region for \HT bin $i$ in the first \MHT and \njets bins.
The \smhtsim term represents a correction for \MHT bin $j$
with respect to the first \MHT bin,
and the \snjets term a correction for \njets bin $k$
with respect to the first \njets bin.
The \kht and \snjets terms are determined from a fit to data
in the $200<\MHT<300\GeV$ bin,
with the non-QCD SM background taken into account.
The \smhtsim terms are taken from the QCD multijet simulation.
Based on studies of the differing contributions of
events in which the jet with the largest \pt mismeasurement
is or is not amongst the four highest \pt jets,
uncertainties of 50, 100, and 100\% are assigned to the
\MHT 300--500, 500--750, and $\geq 750\GeV$ bins,
respectively,
to account for
potential differences between data and simulation in the \smht factors.
Weighted results for \rqcd are calculated
when recombining the \MHT and \njets results
to correspond to the nominal search regions.
Figure~\ref{fig:qcd-mdp-closure} presents
closure test results for the method.

For the lowest \MHT search region,
the uncertainty in the prediction of the QCD multijet
background is dominated by the uncertainties
in \kht and \snjets,
which themselves are mostly due to uncertainties
in the non-QCD SM background in the search regions.
For the two higher \MHT search regions,
the uncertainty in \smhtsim
and the limited statistical precision of the low-\dphii CR
dominate the uncertainty.
The uncertainties related to potential
nonclosure (Fig.~\ref{fig:qcd-mdp-closure})
are either small in comparison
or statistical in nature and are not considered.

\section{Results and interpretation}
\label{sec:results}

\begin{figure*}[ht]
\centering
\includegraphics[width=\cmsFigWidthDouble]{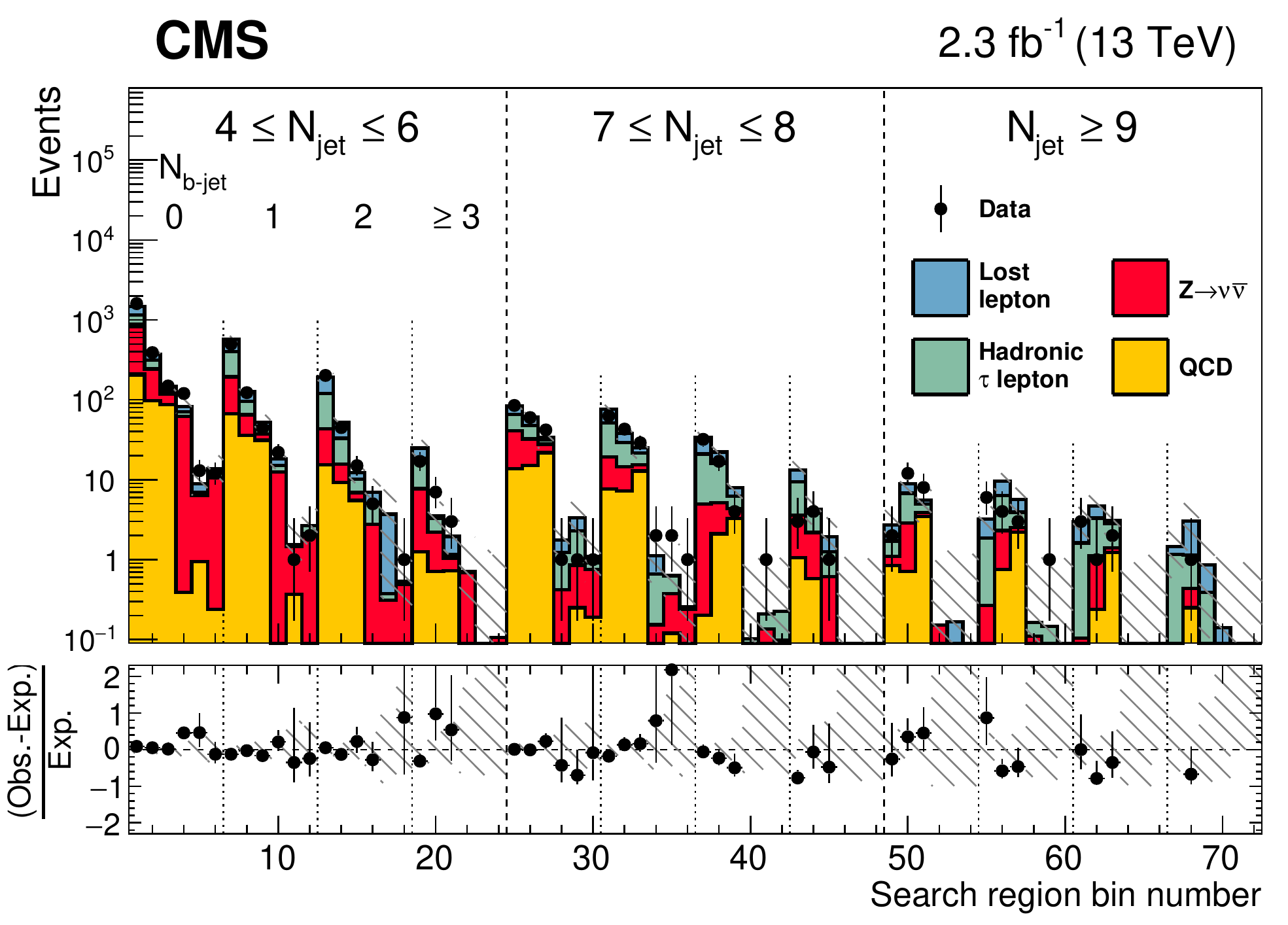}
\caption{
  Observed numbers of events and corresponding prefit
  SM background predictions
  in the 72 search regions of the analysis,
  with fractional differences shown in the lower panel.
  The shaded regions indicate the total uncertainties in the background
  predictions.
  The labeling of the search regions is the same as in Fig.~\ref{fig:lost-lepton-closure}.
}
\label{fig:fit-results}
\end{figure*}

The observed numbers of events in the 72 search regions
are shown in Fig.~\ref{fig:fit-results}
in comparison to the summed predictions for the SM backgrounds,
with numerical values tabulated in \cmsAppendix\ref{sec:prefit}.
The predicted background is observed to be statistically compatible
with the data for all 72 regions.
Therefore, we do not observe evidence for new physics.

\begin{figure*}[htbp]
\centering
\includegraphics[width=0.45\textwidth]{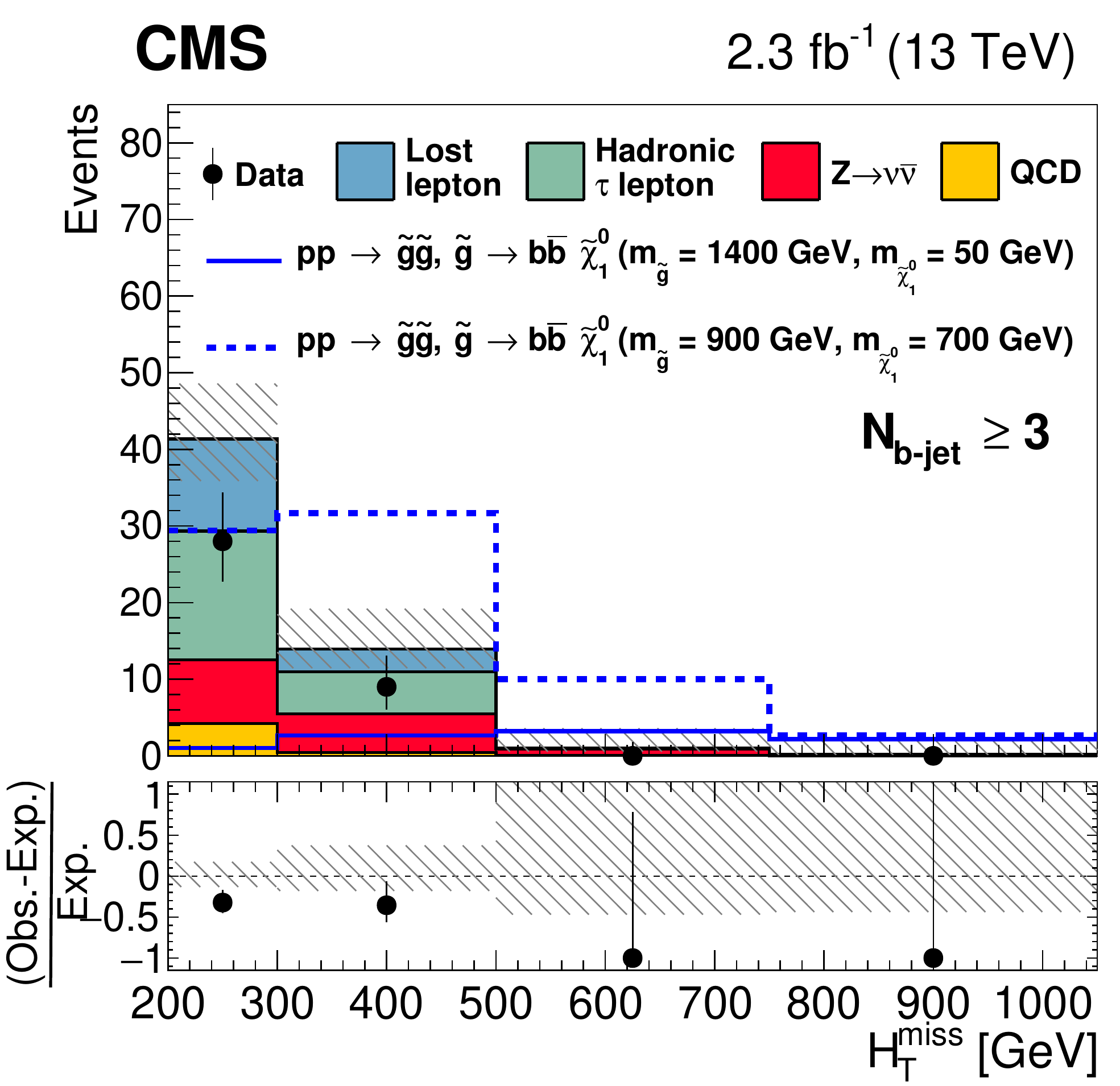}
\includegraphics[width=0.45\textwidth]{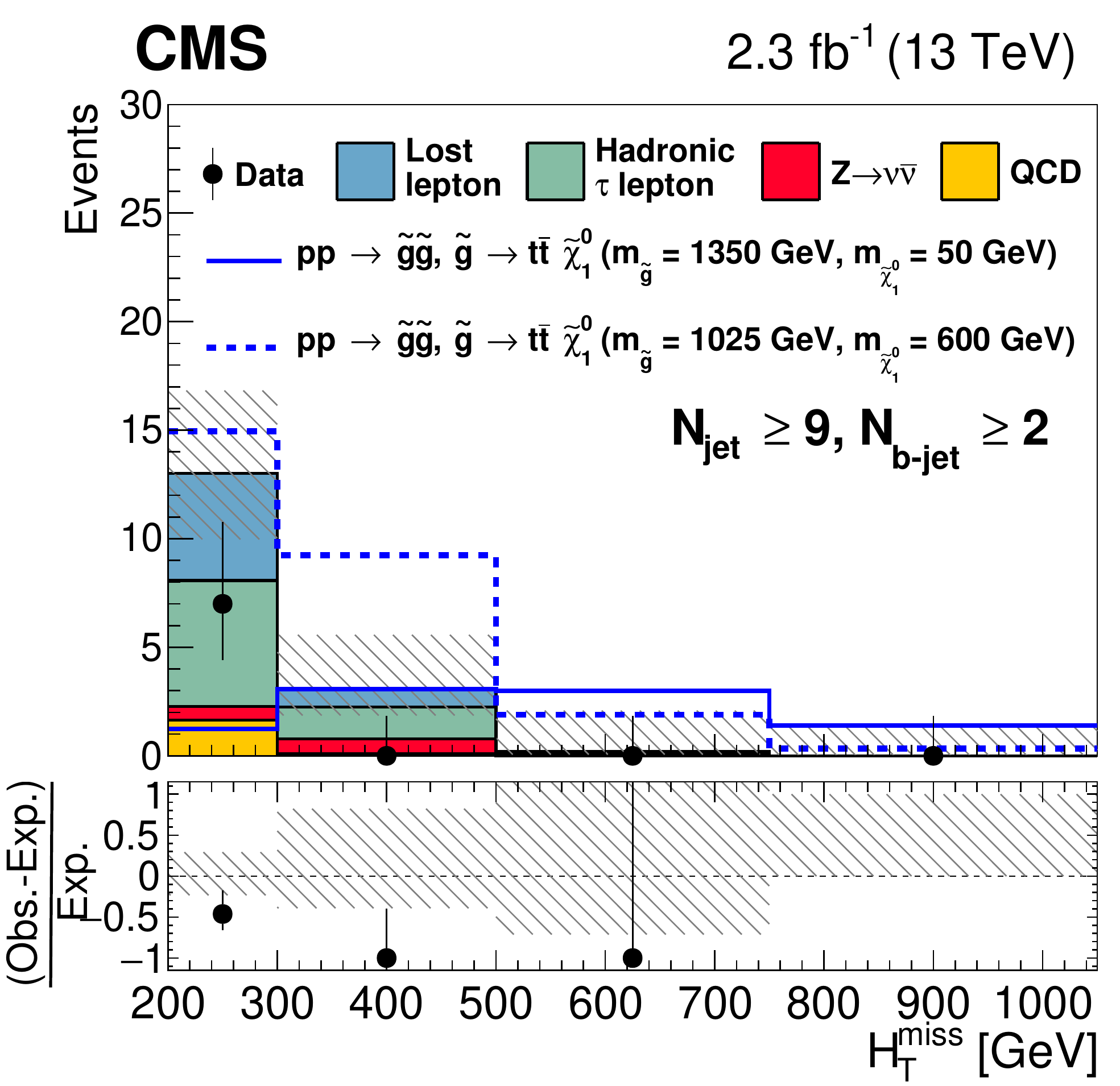}\\
\includegraphics[width=0.45\textwidth]{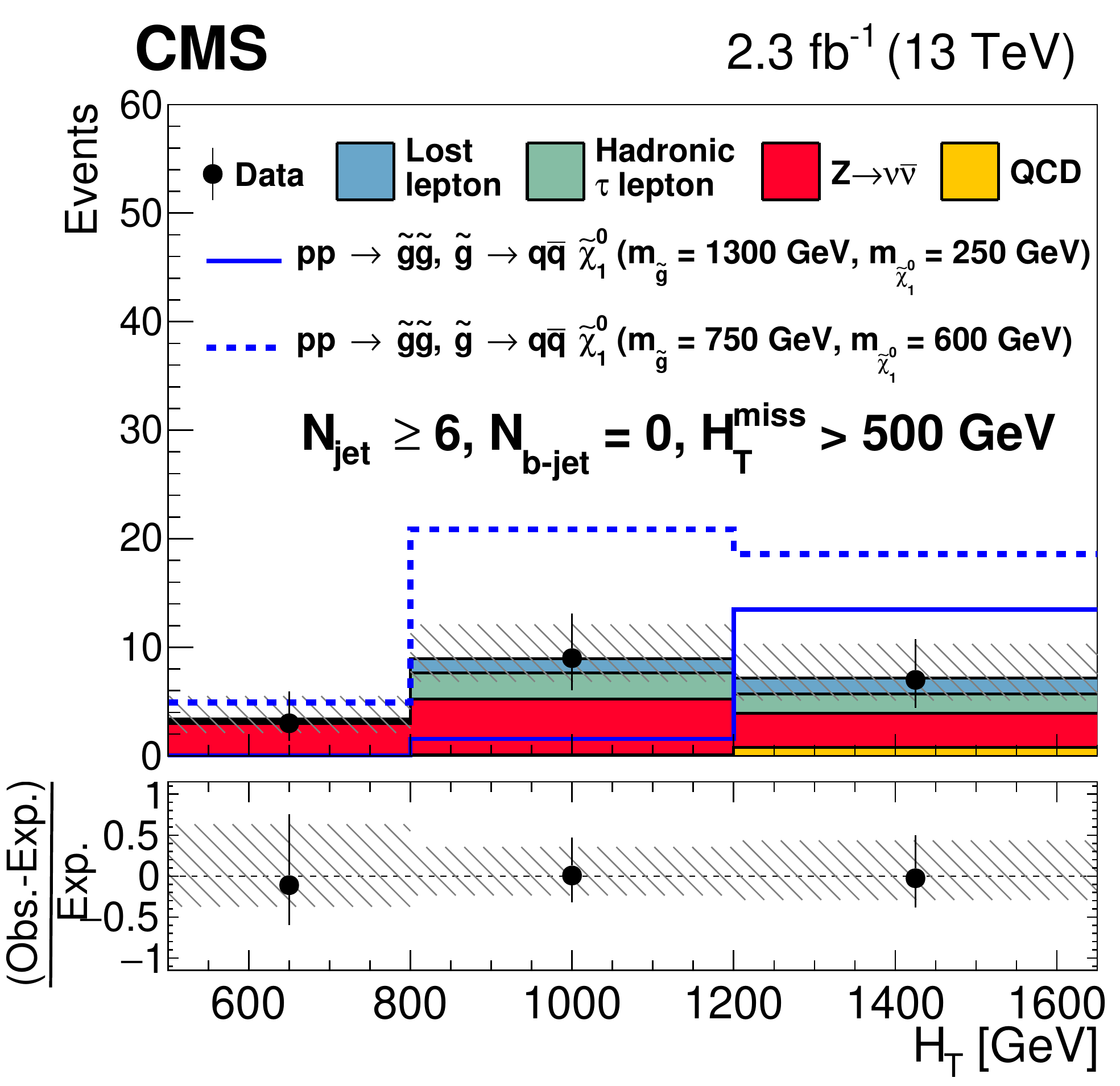}
\includegraphics[width=0.45\textwidth]{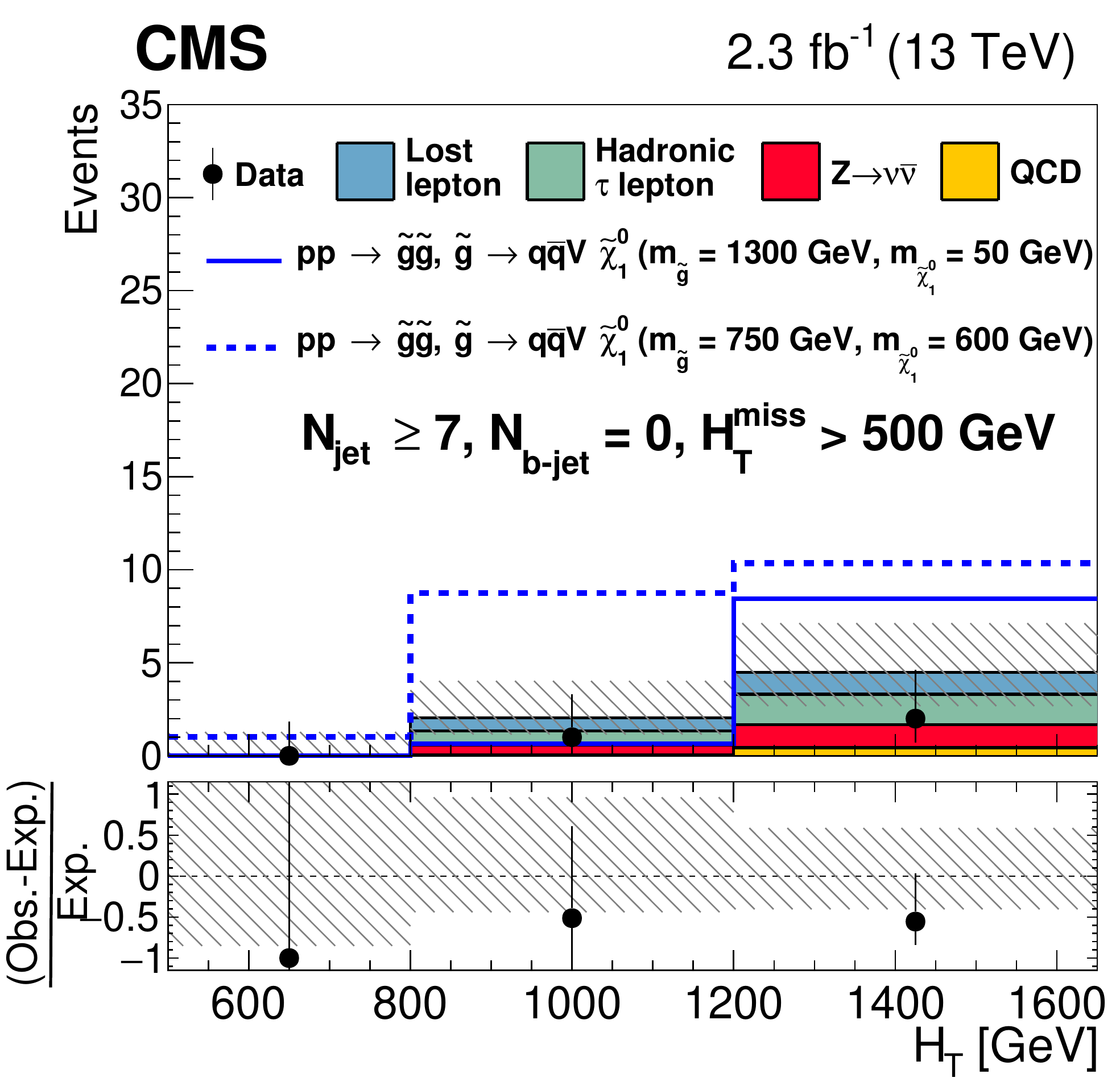}
\caption{
  Observed numbers of events and corresponding SM background predictions
  for intervals of the search region parameter space
  particularly sensitive to the
  (upper left) T1bbbb,
  (upper right) T1tttt,
  (lower left) T1qqqq, and
  (lower right) T5qqqqVV scenarios.
  The selection requirements are given in the figure legends.
  The hatched regions indicate the total uncertainties in the background
  predictions.
  The (unstacked) results for two example signal scenarios are shown in each instance,
  one with ${\mgluino}\gg\mlsp$ and the other with ${\mlsp}\sim\mgluino$.
  Note that for purposes of presentation,
  the four-bin scheme
  discussed in Section~\ref{sec:qcd} is used for the \MHT variable.
  For the T1tttt model,
  the rightmost bin contains both zero predicted background events
  and zero observed events.
}
\label{fig:projections}
\end{figure*}

Figure~\ref{fig:projections} presents
one-dimensional projections of the results in \MHT or \HT
after criteria are imposed,
as indicated in the legends,
to select intervals of the search region parameter space
particularly sensitive to the T1bbbb, T1tttt, T1qqqq, or T5qqqqVV scenario.
In each case,
example distributions are shown for two signal scenarios
not excluded by our Run~1
studies~\cite{Chatrchyan:2013wxa,Chatrchyan:2014lfa}.
These scenarios,
one with ${\mgluino}\gg\mlsp$
and one with ${\mlsp}\sim\mgluino$,
lie well within the parameter space excluded by the present
analysis (see below).

A likelihood fit to data is used to set limits on
the production cross sections of the signal scenarios.
The fitted parameters are the SUSY signal strength,
the yields of the four background classes indicated in Fig.~\ref{fig:fit-results},
and various nuisance parameters.
The limits are determined as a function of \mlsp and \mgluino.
The likelihood function is the product of Poisson probability density functions,
one for each search region,
and constraint terms that account for
uncertainties in the background predictions and signal yields.
These uncertainties are treated as nuisance parameters
with log-normal probability density functions.
Correlations are taken into account where appropriate.
The signal model uncertainties associated with
the renormalization and factorization scales, ISR,
the jet energy scale,
the {\cPqb} jet tagging,
and the statistical fluctuations
vary substantially with the event kinematics
and are evaluated as a function of \mlsp and~\mgluino.
The test statistic is
$q_\mu =  - 2 \ln \left( \mathcal{L}_\mu/\mathcal{L}_\text{max} \right)$,
where $\mathcal{L}_\text{max}$ is the maximum likelihood
determined by allowing all parameters including the
SUSY signal strength $\mu$ to vary,
and $\mathcal{L}_\mu$ is the maximum likelihood for a fixed signal strength.
To set limits,
we use asymptotic results for the test statistic~\cite{Cowan:2010js}
and the CL$_\mathrm{s}$
method described in Refs.~\cite{Junk1999,bib-cls}.
More details are provided in Refs.~\cite{cms-note-2011-005,Khachatryan:2015vra}.

\begin{figure*}[htbp]
\centering
    \includegraphics[width=0.48\textwidth]{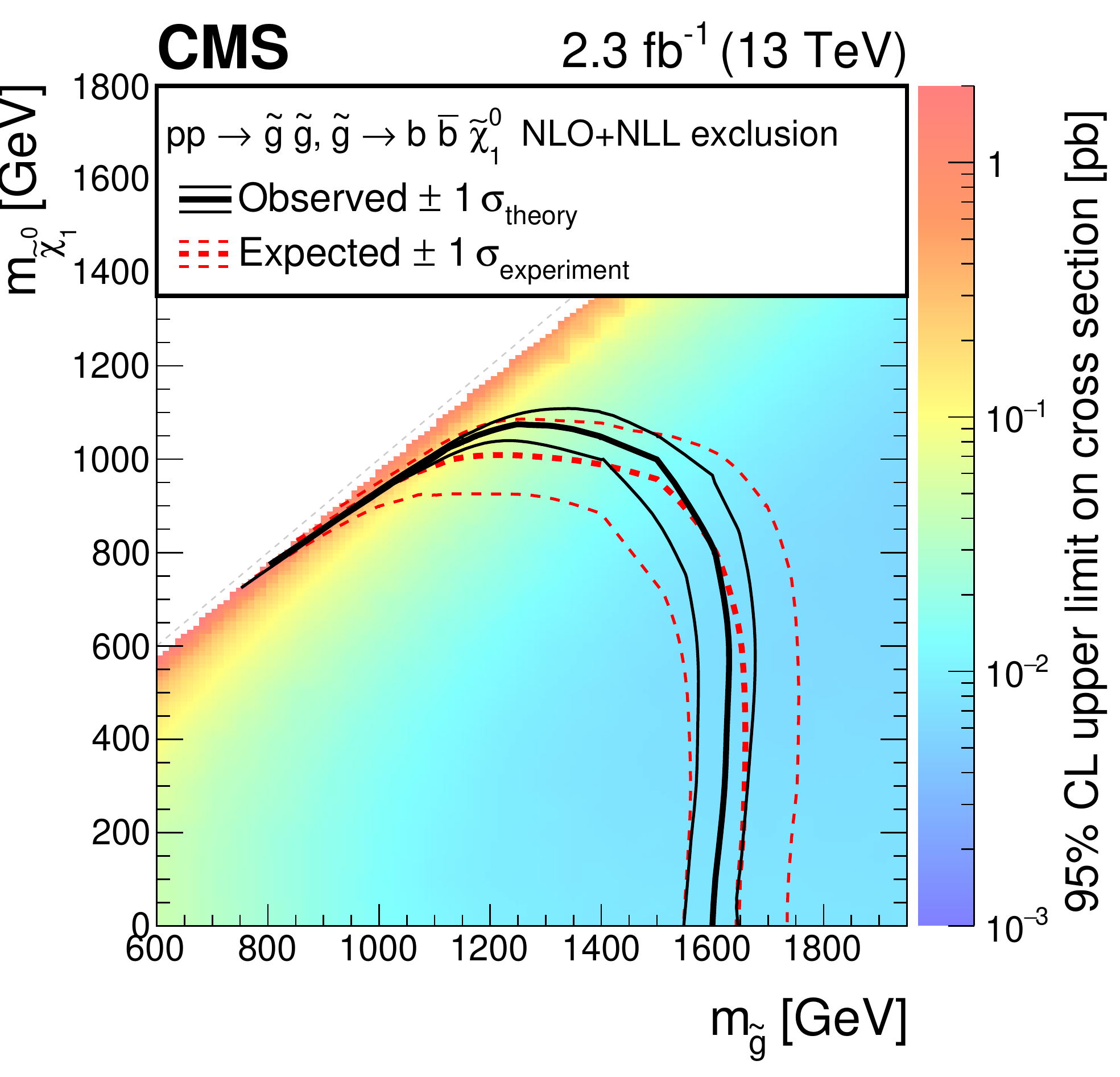}
    \includegraphics[width=0.48\textwidth]{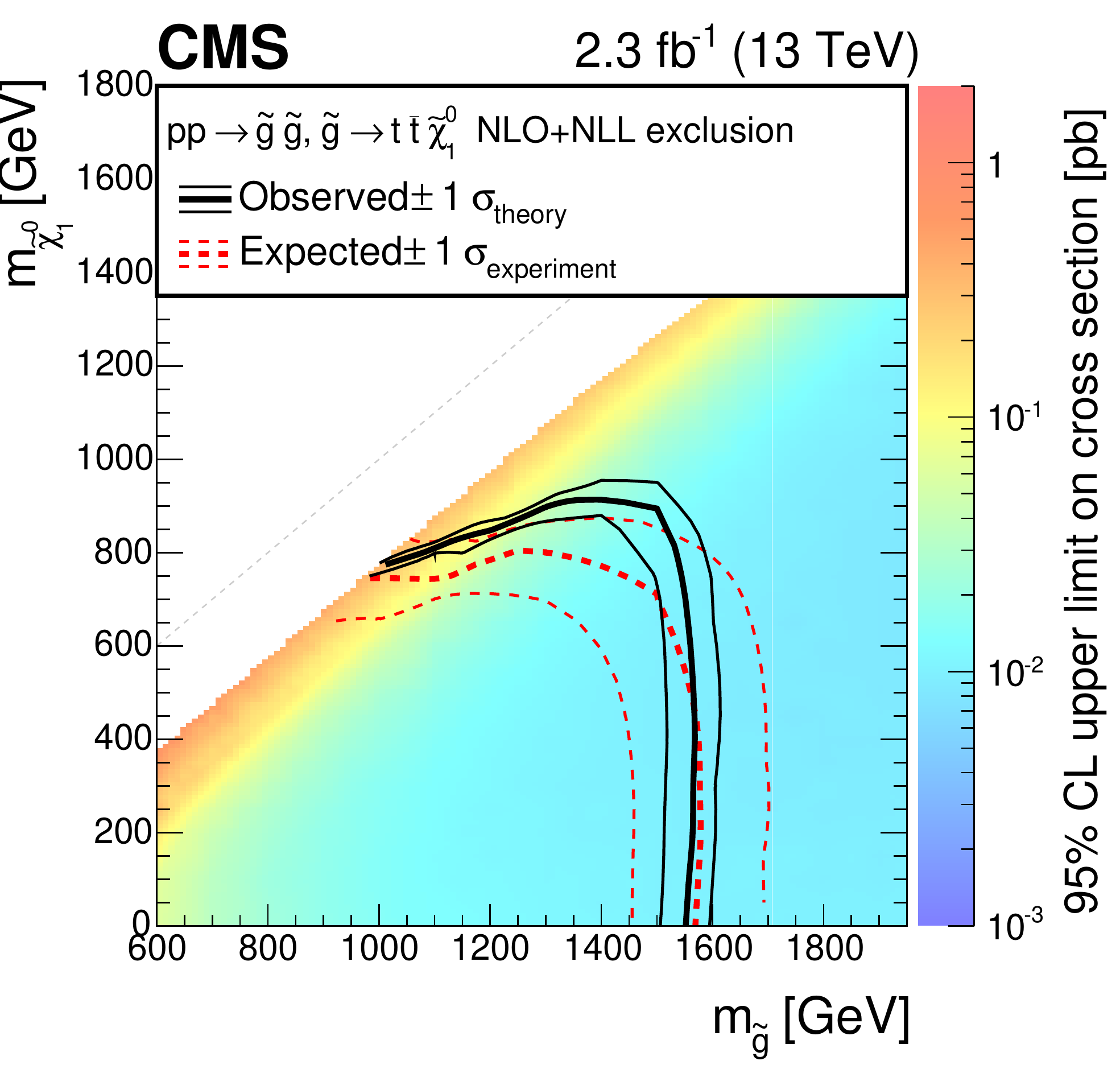} \\
    \includegraphics[width=0.48\textwidth]{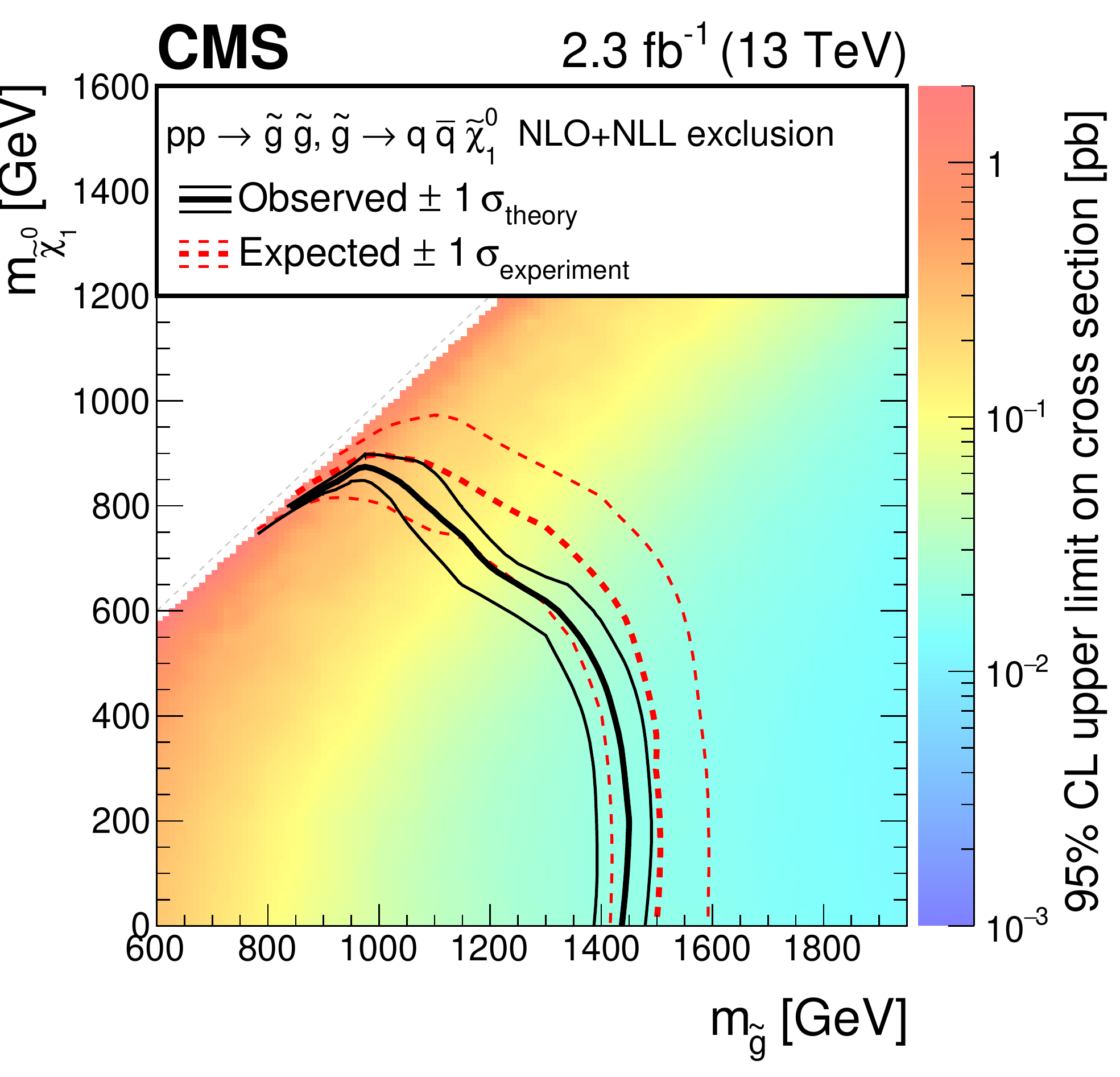}
    \includegraphics[width=0.48\textwidth]{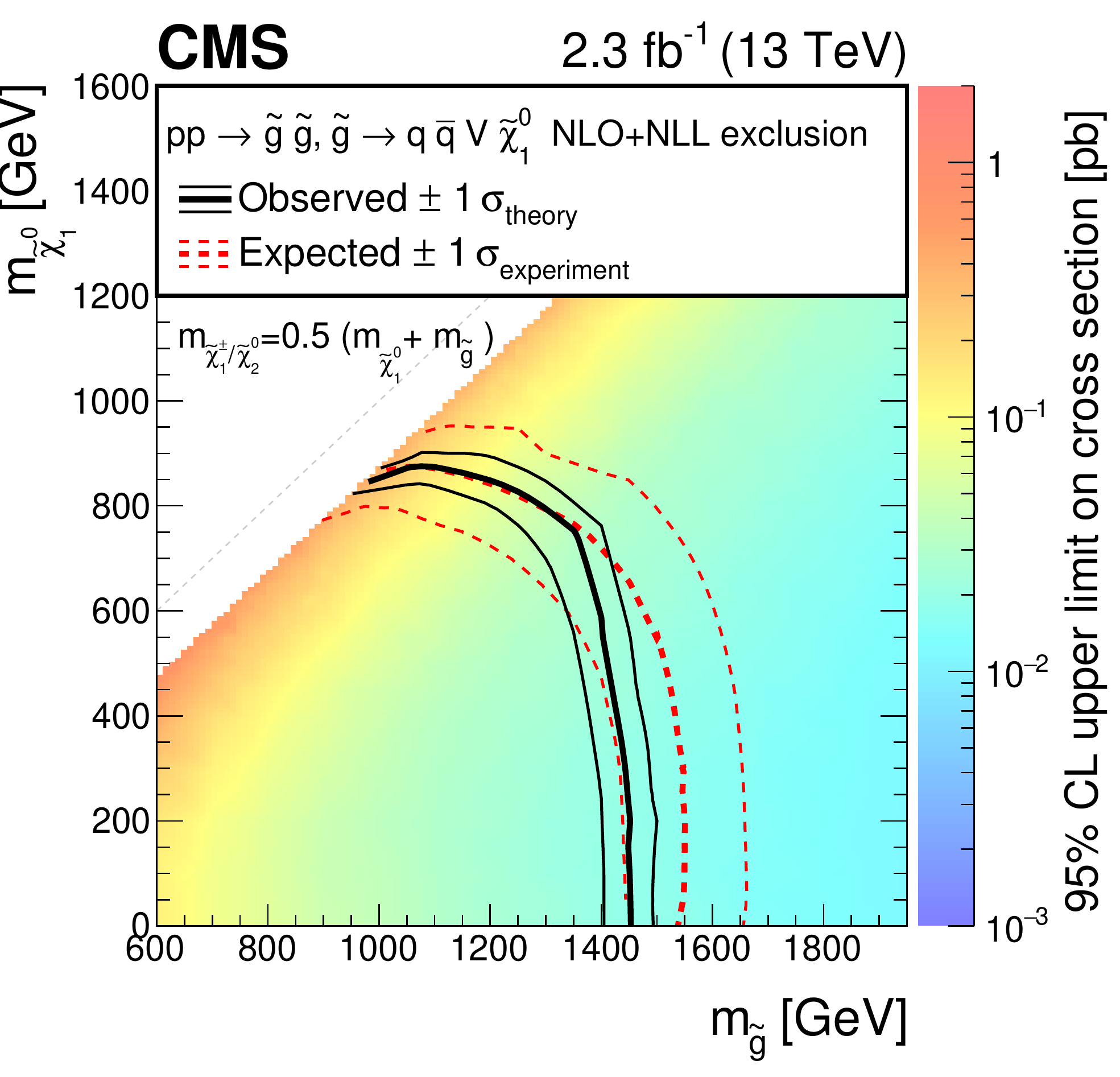}
    \caption{
      The 95\% CL upper limits on the production
      cross sections for the (upper left) T1bbbb,
      (upper right) T1tttt, (lower left) T1qqqq,
      and (lower right) T5qqqqVV simplified models of supersymmetry,
      shown as a function of the gluino and LSP masses \mgluino and~\mlsp.
      For the T5qqqqVV model,
      the masses of the intermediate \PSGczDt and $\PSGcpm_1$ states
      are taken to be the mean of \mlsp and $\mgluino$.
      The solid (black) curves show the observed exclusion contours
      assuming the NLO+NLL cross
      sections~\cite{bib-nlo-nll-01,bib-nlo-nll-02,bib-nlo-nll-03,
      bib-nlo-nll-04,bib-nlo-nll-05},
      with the corresponding $\pm$1~standard
      deviation uncertainties~\cite{Borschensky:2014cia}.
      The dashed (red) curves present the expected limits
      with $\pm$1 standard deviation experimental uncertainties.
      The dashed (grey) lines indicate the $\mlsp=\mgluino$ diagonal.
    }
    \label{fig:limits}
\end{figure*}

We proceed to evaluate 95\% confidence level (CL) upper limits
on the signal cross sections.
The NLO+NLL cross section is used as a reference
to evaluate corresponding 95\% CL exclusion curves.
In addition to the observed limits,
expected limits are derived by evaluating the
expected Poisson fluctuations around the predicted
numbers of background events when evaluating the test statistic.
The potential contributions of signal events to the control regions
are taken into account.
Specifically,
the number of events in each CR is corrected
to include the predicted number of signal events,
in the context of the model being examined,
to derive the total effective number of background events
expected in each search region.
This total effective background is used when determining the limits.

The results are shown in Fig.~\ref{fig:limits}.
For a massless LSP,
we exclude gluinos with masses below 1600, 1550, 1440, and 1450\GeV,
respectively,
for the T1bbbb, T1tttt, T1qqqq, and T5qqqqVV scenarios.
These results significantly extend those we
obtained at $\sqrt{s}=8\TeV$,
for which the corresponding limits are around
1150\GeV~\cite{Chatrchyan:2013wxa,Chatrchyan:2014lfa} for the
three T1 models and 1280\GeV~\cite{Chatrchyan:2014lfa} for the T5 model.

\section{Summary}
\label{sec:summary}

A search is presented for an anomalously high rate of events
with four or more jets,
no identified isolated electron or muon
or isolated charged track,
large scalar sum \HT of jet transverse momenta,
and large missing transverse momentum,
where this latter quantity is measured with the variable \MHT,
the magnitude of the vector sum of jet transverse momenta.
The search is based on a sample of proton-proton collision data
collected at $\sqrt{s}=13\TeV$ with the CMS detector at the CERN LHC in 2015,
corresponding to an integrated luminosity of 2.3\fbinv.
The principal standard model backgrounds,
from events with top quarks,
{\PW} bosons and jets,
{\cPZ} bosons and jets,
and QCD multijet production,
are evaluated using control samples in the data.
The study is performed in the framework of a global likelihood fit
in which the observed numbers of events in 72 exclusive bins
in a four-dimensional array of \MHT,
the number of jets,
the number of tagged bottom quark jets,
and \HT,
are compared to the standard model predictions.
The standard model background estimates are found to agree
with the observed numbers of events within the uncertainties.
The results are interpreted with simplified models that,
in the context of supersymmetry,
correspond to gluino pair production
followed by the decay of each gluino to an undetected
lightest-supersymmetric-particle (LSP) neutralino \PSGczDo
and to a bottom quark-antiquark pair (T1bbbb model),
a top quark-antiquark pair (T1tttt model),
or a light-flavored quark-antiquark pair (T1qqqq model).
We also consider a scenario corresponding to gluino pair
production followed by the decay of each gluino to
a light-flavored quark-antiquark pair and to either
a next-to-lightest neutralino \PSGczDt
or a lightest chargino $\PSGcpm_1$,
with $\PSGczDt\to \Z\PSGczDo$
or $\PSGcpm_1\to\PW^\pm\PSGczDo$
(T5qqqqVV model).
Using the NLO+NLL production cross section as a reference,
and for a massless LSP,
we exclude gluinos with masses below 1600, 1550, 1440, and 1450\GeV
for the four scenarios, respectively,
significantly extending the limits from previous searches.

\begin{acknowledgments}
We congratulate our colleagues in the CERN accelerator departments for the excellent performance of the LHC and thank the technical and administrative staffs at CERN and at other CMS institutes for their contributions to the success of the CMS effort. In addition, we gratefully acknowledge the computing centers and personnel of the Worldwide LHC Computing Grid for delivering so effectively the computing infrastructure essential to our analyses. Finally, we acknowledge the enduring support for the construction and operation of the LHC and the CMS detector provided by the following funding agencies: BMWFW and FWF (Austria); FNRS and FWO (Belgium); CNPq, CAPES, FAPERJ, and FAPESP (Brazil); MES (Bulgaria); CERN; CAS, MoST, and NSFC (China); COLCIENCIAS (Colombia); MSES and CSF (Croatia); RPF (Cyprus); MoER, ERC IUT and ERDF (Estonia); Academy of Finland, MEC, and HIP (Finland); CEA and CNRS/IN2P3 (France); BMBF, DFG, and HGF (Germany); GSRT (Greece); OTKA and NIH (Hungary); DAE and DST (India); IPM (Iran); SFI (Ireland); INFN (Italy); MSIP and NRF (Republic of Korea); LAS (Lithuania); MOE and UM (Malaysia); CINVESTAV, CONACYT, SEP, and UASLP-FAI (Mexico); MBIE (New Zealand); PAEC (Pakistan); MSHE and NSC (Poland); FCT (Portugal); JINR (Dubna); MON, RosAtom, RAS and RFBR (Russia); MESTD (Serbia); SEIDI and CPAN (Spain); Swiss Funding Agencies (Switzerland); MST (Taipei); ThEPCenter, IPST, STAR and NSTDA (Thailand); TUBITAK and TAEK (Turkey); NASU and SFFR (Ukraine); STFC (United Kingdom); DOE and NSF (USA).

Individuals have received support from the Marie-Curie program and the European Research Council and EPLANET (European Union); the Leventis Foundation; the A. P. Sloan Foundation; the Alexander von Humboldt Foundation; the Belgian Federal Science Policy Office; the Fonds pour la Formation \`a la Recherche dans l'Industrie et dans l'Agriculture (FRIA-Belgium); the Agentschap voor Innovatie door Wetenschap en Technologie (IWT-Belgium); the Ministry of Education, Youth and Sports (MEYS) of the Czech Republic; the Council of Science and Industrial Research, India; the HOMING PLUS program of the Foundation for Polish Science, cofinanced from European Union, Regional Development Fund; the OPUS program of the National Science Center (Poland); the Compagnia di San Paolo (Torino); MIUR project 20108T4XTM (Italy); the Thalis and Aristeia programmes cofinanced by EU-ESF and the Greek NSRF; the National Priorities Research Program by Qatar National Research Fund; the Rachadapisek Sompot Fund for Postdoctoral Fellowship, Chulalongkorn University (Thailand); the Chulalongkorn Academic into Its 2nd Century Project Advancement Project (Thailand); and the Welch Foundation, contract C-1845.
\end{acknowledgments}

\clearpage
\bibliography{auto_generated}

\providecommand{\href}[2]{#2}\begingroup\raggedright\begin{thebibliography}{10}%
\makeatletter
\providecommand{\hrefCMSnoop }[0]{\@secondoftwo}%
\makeatother
\providecommand{\doi}{\texttt{doi:}\begingroup \urlstyle{tt}\Url}

\bibitem{Ramond:1971gb}
\hrefCMSnoop {}{P.~Ramond, ``{Dual theory for free fermions}'',} \textit{ Phys.
  Rev. D} \textbf{ 3} (1971) 2415,
\href{http://dx.doi.org/10.1103/PhysRevD.3.2415}{\doi{10.1103/PhysRevD.3.2415}}.
%%CITATION = PHRVA,D3,2415;%%.

\bibitem{Golfand:1971iw}
\hrefCMSnoop {}{Y.~A. Golfand and E.~P. Likhtman, ``{Extension of the algebra
  of {P}oincar\'{e} group generators and violation of {P} invariance}'',}
  \textit{ JETP Lett.} \textbf{ 13} (1971)
323.
%%CITATION = JTPLA,13,323;%%.

\bibitem{Neveu:1971rx}
\hrefCMSnoop {}{A.~Neveu and J.~H. Schwarz, ``{Factorizable dual model of
  pions}'',} \textit{ Nucl. Phys. B} \textbf{ 31} (1971) 86,
\href{http://dx.doi.org/10.1016/0550-3213(71)90448-2}{\doi{10.1016/0550-3213(71)90448-2}}.
%%CITATION = NUPHA,B31,86;%%.

\bibitem{Volkov:1972jx}
\hrefCMSnoop {}{D.~V. Volkov and V.~P. Akulov, ``{Possible universal neutrino
  interaction}'',} \textit{ JETP Lett.} \textbf{ 16} (1972)
438.
%%CITATION = JTPLA,16,438;%%.

\bibitem{Wess:1973kz}
\hrefCMSnoop {}{J.~Wess and B.~Zumino, ``{A {L}agrangian model invariant under
  supergauge transformations}'',} \textit{ Phys. Lett. B} \textbf{ 49} (1974)
  52,
\href{http://dx.doi.org/10.1016/0370-2693(74)90578-4}{\doi{10.1016/0370-2693(74)90578-4}}.
%%CITATION = PHLTA,B49,52;%%.

\bibitem{Wess:1974tw}
\hrefCMSnoop {}{J.~Wess and B.~Zumino, ``{Supergauge transformations in four
  dimensions}'',} \textit{ Nucl. Phys. B} \textbf{ 70} (1974) 39,
\href{http://dx.doi.org/10.1016/0550-3213(74)90355-1}{\doi{10.1016/0550-3213(74)90355-1}}.
%%CITATION = NUPHA,B70,39;%%.

\bibitem{Fayet:1974pd}
\hrefCMSnoop {}{P.~Fayet, ``{Supergauge invariant extension of the {H}iggs
  mechanism and a model for the electron and its neutrino}'',} \textit{ Nucl.
  Phys. B} \textbf{ 90} (1975) 104,
\href{http://dx.doi.org/10.1016/0550-3213(75)90636-7}{\doi{10.1016/0550-3213(75)90636-7}}.
%%CITATION = NUPHA,B90,104;%%.

\bibitem{Nilles:1983ge}
\hrefCMSnoop {}{H.~P. Nilles, ``{Supersymmetry, supergravity and particle
  physics}'',} \textit{ Phys. Rep.} \textbf{ 110} (1984) 1,
\href{http://dx.doi.org/10.1016/0370-1573(84)90008-5}{\doi{10.1016/0370-1573(84)90008-5}}.
%%CITATION = PRPLC,110,1;%%.

\bibitem{Barbieri:1987fn}
\hrefCMSnoop {}{R.~Barbieri and G.~F. Giudice, ``{Upper Bounds on
  Supersymmetric Particle Masses}'',} \textit{ Nucl. Phys. B} \textbf{ 306}
  (1988) 63,
\href{http://dx.doi.org/10.1016/0550-3213(88)90171-X}{\doi{10.1016/0550-3213(88)90171-X}}.
%%CITATION = NUPHA,B306,63;%%.

\bibitem{Dimopoulos:1995mi}
\hrefCMSnoop {}{S.~Dimopoulos and G.~F. Giudice, ``{Naturalness constraints in
  supersymmetric theories with nonuniversal soft terms}'',} \textit{ Phys.
  Lett. B} \textbf{ 357} (1995) 573,
  \href{http://dx.doi.org/10.1016/0370-2693(95)00961-J}{\doi{10.1016/0370-2693(95)00961-J}},
\href{http://www.arXiv.org/abs/hep-ph/9507282}{\texttt{arXiv:hep-ph/9507282}}.
%%CITATION = HEP-PH/9507282;%%.

\bibitem{Barbieri:2009ev}
\hrefCMSnoop {}{R.~Barbieri and D.~Pappadopulo, ``{S-particles at their
  naturalness limits}'',} \textit{ JHEP} \textbf{ 10} (2009) 061,
  \href{http://dx.doi.org/10.1088/1126-6708/2009/10/061}{\doi{10.1088/1126-6708/2009/10/061}},
\href{http://www.arXiv.org/abs/0906.4546}{\texttt{arXiv:0906.4546}}.
%%CITATION = ARXIV:0906.4546;%%.

\bibitem{Papucci:2011wy}
\hrefCMSnoop {}{M.~Papucci, J.~T. Ruderman, and A.~Weiler, ``{Natural {SUSY}
  endures}'',} \textit{ JHEP} \textbf{ 09} (2012) 035,
  \href{http://dx.doi.org/10.1007/JHEP09(2012)035}{\doi{10.1007/JHEP09(2012)035}},
\href{http://www.arXiv.org/abs/1110.6926}{\texttt{arXiv:1110.6926}}.
%%CITATION = ARXIV:1110.6926;%%.

\bibitem{bib-rparity}
\hrefCMSnoop {}{G.~R. Farrar and P.~Fayet, ``Phenomenology of the production,
  decay, and detection of new hadronic states associated with supersymmetry'',}
  \textit{ Phys. Lett. B} \textbf{ 76} (1978) 575,
  \href{http://dx.doi.org/10.1016/0370-2693(78)90858-4}{\doi{10.1016/0370-2693(78)90858-4}}.

\bibitem{Aad:2015iea}
\hrefCMSnoop {}{{ATLAS Collaboration}, ``{Summary of the searches for squarks
  and gluinos using $\sqrt{s}=8$~TeV pp collisions with the ATLAS experiment at
  the LHC}'',} \textit{ JHEP} \textbf{ 10} (2015) 054,
  \href{http://dx.doi.org/10.1007/JHEP10(2015)054}{\doi{10.1007/JHEP10(2015)054}},
\href{http://www.arXiv.org/abs/1507.05525}{\texttt{arXiv:1507.05525}}.
%%CITATION = ARXIV:1507.05525;%%.

\bibitem{Khachatryan:2015vra}
\hrefCMSnoop {}{{CMS Collaboration}, ``{Searches for supersymmetry using the
  M$_{T2}$ variable in hadronic events produced in pp collisions at 8~TeV}'',}
  \textit{ JHEP} \textbf{ 05} (2015) 078,
  \href{http://dx.doi.org/10.1007/JHEP05(2015)078}{\doi{10.1007/JHEP05(2015)078}},
\href{http://www.arXiv.org/abs/1502.04358}{\texttt{arXiv:1502.04358}}.
%%CITATION = ARXIV:1502.04358;%%.

\bibitem{Khachatryan:2015pwa}
\hrefCMSnoop {}{{CMS Collaboration}, ``{Search for supersymmetry using razor
  variables in events with b-tagged jets in pp collisions at $\sqrt{s}
  =8$~TeV}'',} \textit{ Phys. Rev. D} \textbf{ 91} (2015) 052018,
  \href{http://dx.doi.org/10.1103/PhysRevD.91.052018}{\doi{10.1103/PhysRevD.91.052018}},
\href{http://www.arXiv.org/abs/1502.00300}{\texttt{arXiv:1502.00300}}.
%%CITATION = ARXIV:1502.00300;%%.

\bibitem{bib-sms-1}
N.~Arkani-Hamed\hrefCMSnoop {}{ {et~al.}, ``{{MARMOSET}: The path from {LHC}
  data to the new standard model via on-shell effective theories}'',} (2007).
\href{http://www.arXiv.org/abs/hep-ph/0703088}{\texttt{arXiv:hep-ph/0703088}}.
%%CITATION = HEP-PH/0703088;%%.

\bibitem{bib-sms-2}
\hrefCMSnoop {}{J.~Alwall, P.~Schuster, and N.~Toro, ``Simplified models for a
  first characterization of new physics at the {LHC}'',} \textit{ Phys. Rev. D}
  \textbf{ 79} (2009) 075020,
  \href{http://dx.doi.org/10.1103/PhysRevD.79.075020}{\doi{10.1103/PhysRevD.79.075020}},
\href{http://www.arXiv.org/abs/0810.3921}{\texttt{arXiv:0810.3921}}.
%%CITATION = ARXIV:0810.3921;%%.

\bibitem{bib-sms-3}
\hrefCMSnoop {}{J.~Alwall, M.-P. Le, M.~Lisanti, and J.~G. Wacker,
  ``{Model-independent jets plus missing energy searches}'',} \textit{ Phys.
  Rev. D} \textbf{ 79} (2009) 015005,
  \href{http://dx.doi.org/10.1103/PhysRevD.79.015005}{\doi{10.1103/PhysRevD.79.015005}},
\href{http://www.arXiv.org/abs/0809.3264}{\texttt{arXiv:0809.3264}}.
%%CITATION = ARXIV:0809.3264;%%.

\bibitem{bib-sms-4}
D.~Alves\hrefCMSnoop {}{ {et~al.}, ``Simplified models for {LHC} new physics
  searches'',} \textit{ J. Phys. G} \textbf{ 39} (2012) 105005,
  \href{http://dx.doi.org/10.1088/0954-3899/39/10/105005}{\doi{10.1088/0954-3899/39/10/105005}},
\href{http://www.arXiv.org/abs/1105.2838}{\texttt{arXiv:1105.2838}}.
%%CITATION = ARXIV:1105.2838;%%.

\bibitem{Chatrchyan:2013sza}
\hrefCMSnoop {}{{CMS Collaboration}, ``{Interpretation of searches for
  supersymmetry with simplified models}'',} \textit{ Phys. Rev. D} \textbf{ 88}
  (2013) 052017,
  \href{http://dx.doi.org/10.1103/PhysRevD.88.052017}{\doi{10.1103/PhysRevD.88.052017}},
\href{http://www.arXiv.org/abs/1301.2175}{\texttt{arXiv:1301.2175}}.
%%CITATION = ARXIV:1301.2175;%%.

\bibitem{Chatrchyan:2013wxa}
\hrefCMSnoop {}{{CMS Collaboration}, ``{Search for gluino mediated bottom- and
  top-squark production in multijet final states in pp collisions at 8~TeV}'',}
  \textit{ Phys. Lett. B} \textbf{ 725} (2013) 243,
  \href{http://dx.doi.org/10.1016/j.physletb.2013.06.058}{\doi{10.1016/j.physletb.2013.06.058}},
\href{http://www.arXiv.org/abs/1305.2390}{\texttt{arXiv:1305.2390}}.
%%CITATION = ARXIV:1305.2390;%%.

\bibitem{Chatrchyan:2014lfa}
\hrefCMSnoop {}{{CMS Collaboration}, ``{Search for new physics in the multijet
  and missing transverse momentum final state in proton-proton collisions at
  $\sqrt{s}= 8$ TeV}'',} \textit{ JHEP} \textbf{ 06} (2014) 055,
  \href{http://dx.doi.org/10.1007/JHEP06(2014)055}{\doi{10.1007/JHEP06(2014)055}},
\href{http://www.arXiv.org/abs/1402.4770}{\texttt{arXiv:1402.4770}}.
%%CITATION = ARXIV:1402.4770;%%.

\bibitem{Chatrchyan:2008aa}
\hrefCMSnoop {}{{CMS Collaboration}, ``{The {CMS} experiment at the {CERN}
  {LHC}}'',} \textit{ JINST} \textbf{ 3} (2008) S08004,
\href{http://dx.doi.org/10.1088/1748-0221/3/08/S08004}{\doi{10.1088/1748-0221/3/08/S08004}}.
%%CITATION = JINST,3,S08004;%%.

\bibitem{cms-pas-pft-09-001}
\href {http://cdsweb.cern.ch/record/1194487}{{CMS Collaboration}, ``Particle
  flow event reconstruction in {CMS} and performance for jets, taus
  and~\MET'',} CMS Physics Analysis Summary CMS-PAS-PFT-09-001, CERN, 2009.

\bibitem{cms-pas-pft-10-001}
\href {http://cdsweb.cern.ch/record/1247373}{{CMS Collaboration},
  ``Commissioning of the particle-flow event reconstruction with the first LHC
  collisions recorded in the CMS detector'',} CMS Physics Analysis Summary
  CMS-PAS-PFT-10-001, CERN, 2010.

\bibitem{Khachatryan:2015hwa}
\hrefCMSnoop {}{{CMS Collaboration}, ``{Performance of electron reconstruction
  and selection with the CMS detector in proton-proton collisions at $\sqrt{s}=
  8\TeV$}'',} \textit{ JINST} \textbf{ 10} (2015) P06005,
  \href{http://dx.doi.org/10.1088/1748-0221/10/06/P06005}{\doi{10.1088/1748-0221/10/06/P06005}},
\href{http://www.arXiv.org/abs/1502.02701}{\texttt{arXiv:1502.02701}}.
%%CITATION = ARXIV:1502.02701;%%.

\bibitem{Chatrchyan:2013sba}
\hrefCMSnoop {}{{CMS Collaboration}, ``{The performance of the CMS muon
  detector in proton-proton collisions at $\sqrt{s} = 7\TeV$ at the LHC}'',}
  \textit{ JINST} \textbf{ 8} (2013) P11002,
  \href{http://dx.doi.org/10.1088/1748-0221/8/11/P11002}{\doi{10.1088/1748-0221/8/11/P11002}},
\href{http://www.arXiv.org/abs/1306.6905}{\texttt{arXiv:1306.6905}}.
%%CITATION = ARXIV:1306.6905;%%.

\bibitem{CMS-PAS-JME-14-001}
\href {http://cds.cern.ch/record/1751454}{{{CMS}} Collaboration, ``Study of
  pileup removal algorithms for jets'',} CMS Physics Analysis Summary
  CMS-PAS-JME-14-001, CERN, 2014.

\bibitem{Cacciari:2008gp}
\hrefCMSnoop {}{M.~Cacciari, G.~P. Salam, and G.~Soyez, ``{The Anti-$k_t$ jet
  clustering algorithm}'',} \textit{ JHEP} \textbf{ 04} (2008) 063,
  \href{http://dx.doi.org/10.1088/1126-6708/2008/04/063}{\doi{10.1088/1126-6708/2008/04/063}},
\href{http://www.arXiv.org/abs/0802.1189}{\texttt{arXiv:0802.1189}}.
%%CITATION = ARXIV:0802.1189;%%.

\bibitem{Cacciari:2011ma}
\hrefCMSnoop {}{M.~Cacciari, G.~P. Salam, and G.~Soyez, ``{FastJet user
  manual}'',} \textit{ Eur. Phys. J. C} \textbf{ 72} (2012) 1896,
  \href{http://dx.doi.org/10.1140/epjc/s10052-012-1896-2}{\doi{10.1140/epjc/s10052-012-1896-2}},
\href{http://www.arXiv.org/abs/1111.6097}{\texttt{arXiv:1111.6097}}.
%%CITATION = ARXIV:1111.6097;%%.

\bibitem{cms-pas-jme-10-003}
\href {http://cdsweb.cern.ch/record/1279362}{{CMS Collaboration}, ``Jet
  performance in pp collisions at $\sqrt{s}=7$~TeV'',} CMS Physics Analysis
  Summary CMS-PAS-JME-10-003, CERN, 2010.

\bibitem{Cacciari:2007fd}
\hrefCMSnoop {}{M.~Cacciari and G.~P. Salam, ``{Pileup subtraction using jet
  areas}'',} \textit{ Phys. Lett. B} \textbf{ 659} (2008) 119,
  \href{http://dx.doi.org/10.1016/j.physletb.2007.09.077}{\doi{10.1016/j.physletb.2007.09.077}},
\href{http://www.arXiv.org/abs/0707.1378}{\texttt{arXiv:0707.1378}}.
%%CITATION = ARXIV:0707.1378;%%.

\bibitem{Chatrchyan:2011ds}
\hrefCMSnoop {}{{CMS Collaboration}, ``{Determination of jet energy calibration
  and transverse momentum resolution in {CMS}}'',} \textit{ JINST} \textbf{ 6}
  (2011) P11002,
  \href{http://dx.doi.org/10.1088/1748-0221/6/11/P11002}{\doi{10.1088/1748-0221/6/11/P11002}},
\href{http://www.arXiv.org/abs/1107.4277}{\texttt{arXiv:1107.4277}}.
%%CITATION = ARXIV:1107.4277;%%.

\bibitem{CMS-PAS-BTV-15-001}
\href {https://cds.cern.ch/record/2138504}{{CMS Collaboration},
  ``Identification of $\cPqb$ quark jets at the CMS experiment in the LHC
  Run~2'',} CMS Physics Analysis Summary CMS-PAS-BTV-15-001, CERN, 2016.

\bibitem{Arnison:1983rp}
\hrefCMSnoop {}{{UA1} Collaboration, ``{Experimental Observation of Isolated
  Large Transverse Energy Electrons with Associated Missing Energy at
  $\sqrt{s}= 540$~GeV}'',} \textit{ Phys. Lett. B} \textbf{ 122} (1983) 103,
\href{http://dx.doi.org/10.1016/0370-2693(83)91177-2}{\doi{10.1016/0370-2693(83)91177-2}}.
%%CITATION = PHLTA,B122,103;%%.

\bibitem{Alwall:2014hca}
J.~Alwall\hrefCMSnoop {}{ {et~al.}, ``{The automated computation of tree-level
  and next-to-leading order differential cross sections, and their matching to
  parton shower simulations}'',} \textit{ JHEP} \textbf{ 07} (2014) 079,
  \href{http://dx.doi.org/10.1007/JHEP07(2014)079}{\doi{10.1007/JHEP07(2014)079}},
\href{http://www.arXiv.org/abs/1405.0301}{\texttt{arXiv:1405.0301}}.
%%CITATION = ARXIV:1405.0301;%%.

\bibitem{Nason:2004rx}
\hrefCMSnoop {}{P.~Nason, ``{A new method for combining NLO QCD with shower
  Monte Carlo algorithms}'',} \textit{ JHEP} \textbf{ 11} (2004) 040,
  \href{http://dx.doi.org/10.1088/1126-6708/2004/11/040}{\doi{10.1088/1126-6708/2004/11/040}},
\href{http://www.arXiv.org/abs/hep-ph/0409146}{\texttt{arXiv:hep-ph/0409146}}.
%%CITATION = HEP-PH/0409146;%%.

\bibitem{Frixione:2007vw}
\hrefCMSnoop {}{S.~Frixione, P.~Nason, and C.~Oleari, ``{Matching NLO QCD
  computations with Parton Shower simulations: the POWHEG method}'',} \textit{
  JHEP} \textbf{ 11} (2007) 070,
  \href{http://dx.doi.org/10.1088/1126-6708/2007/11/070}{\doi{10.1088/1126-6708/2007/11/070}},
\href{http://www.arXiv.org/abs/0709.2092}{\texttt{arXiv:0709.2092}}.
%%CITATION = ARXIV:0709.2092;%%.

\bibitem{Alioli:2010xd}
\hrefCMSnoop {}{S.~Alioli, P.~Nason, C.~Oleari, and E.~Re, ``{A general
  framework for implementing NLO calculations in shower Monte Carlo programs:
  the POWHEG BOX}'',} \textit{ JHEP} \textbf{ 06} (2010) 043,
  \href{http://dx.doi.org/10.1007/JHEP06(2010)043}{\doi{10.1007/JHEP06(2010)043}},
\href{http://www.arXiv.org/abs/1002.2581}{\texttt{arXiv:1002.2581}}.
%%CITATION = ARXIV:1002.2581;%%.

\bibitem{Alioli:2009je}
\hrefCMSnoop {}{S.~Alioli, P.~Nason, C.~Oleari, and E.~Re, ``{NLO single-top
  production matched with shower in POWHEG: $s$- and $t$-channel
  contributions}'',} \textit{ JHEP} \textbf{ 09} (2009) 111,
  \href{http://dx.doi.org/10.1088/1126-6708/2009/09/111}{\doi{10.1088/1126-6708/2009/09/111}},
  \href{http://www.arXiv.org/abs/0907.4076}{\texttt{arXiv:0907.4076}}.
[Erratum: \DOI{10.1007/JHEP02(2010)011}].
%%CITATION = ARXIV:0907.4076;%%.

\bibitem{Re:2010bp}
\hrefCMSnoop {}{E.~Re, ``{Single-top Wt-channel production matched with parton
  showers using the POWHEG method}'',} \textit{ Eur. Phys. J. C} \textbf{ 71}
  (2011) 1547,
  \href{http://dx.doi.org/10.1140/epjc/s10052-011-1547-z}{\doi{10.1140/epjc/s10052-011-1547-z}},
\href{http://www.arXiv.org/abs/1009.2450}{\texttt{arXiv:1009.2450}}.
%%CITATION = ARXIV:1009.2450;%%.

\bibitem{Agostinelli:2002hh}
\hrefCMSnoop {}{{GEANT4} Collaboration, ``{GEANT4}---a simulation toolkit'',}
  \textit{ Nucl. Instrum. Meth. A} \textbf{ 506} (2003) 250,
\href{http://dx.doi.org/10.1016/S0168-9002(03)01368-8}{\doi{10.1016/S0168-9002(03)01368-8}}.
%%CITATION = NUIMA,A506,250;%%.

\bibitem{Melia:2011tj}
\hrefCMSnoop {}{T.~Melia, P.~Nason, R.~Rontsch, and G.~Zanderighi,
  ``{W$^+$W$^-$, WZ and ZZ production in the POWHEG BOX}'',} \textit{ JHEP}
  \textbf{ 11} (2011) 078,
  \href{http://dx.doi.org/10.1007/JHEP11(2011)078}{\doi{10.1007/JHEP11(2011)078}},
\href{http://www.arXiv.org/abs/1107.5051}{\texttt{arXiv:1107.5051}}.
%%CITATION = ARXIV:1107.5051;%%.

\bibitem{Beneke:2011mq}
\hrefCMSnoop {}{M.~Beneke, P.~Falgari, S.~Klein, and C.~Schwinn, ``{Hadronic
  top-quark pair production with NNLL threshold resummation}'',} \textit{ Nucl.
  Phys. B} \textbf{ 855} (2012) 695,
  \href{http://dx.doi.org/10.1016/j.nuclphysb.2011.10.021}{\doi{10.1016/j.nuclphysb.2011.10.021}},
\href{http://www.arXiv.org/abs/1109.1536}{\texttt{arXiv:1109.1536}}.
%%CITATION = ARXIV:1109.1536;%%.

\bibitem{Cacciari:2011hy}
M.~Cacciari\hrefCMSnoop {}{ {et~al.}, ``{Top-pair production at hadron
  colliders with next-to-next-to-leading logarithmic soft-gluon
  resummation}'',} \textit{ Phys. Lett. B} \textbf{ 710} (2012) 612,
  \href{http://dx.doi.org/10.1016/j.physletb.2012.03.013}{\doi{10.1016/j.physletb.2012.03.013}},
\href{http://www.arXiv.org/abs/1111.5869}{\texttt{arXiv:1111.5869}}.
%%CITATION = ARXIV:1111.5869;%%.

\bibitem{Baernreuther:2012ws}
\hrefCMSnoop {}{P.~B{\"{a}}rnreuther, M.~Czakon, and A.~Mitov, ``{Percent Level
  Precision Physics at the Tevatron: First Genuine NNLO QCD Corrections to
  $\PQq\PAQq\to\ttbar + X$}'',} \textit{ Phys. Rev. Lett.} \textbf{ 109} (2012)
  132001,
  \href{http://dx.doi.org/10.1103/PhysRevLett.109.132001}{\doi{10.1103/PhysRevLett.109.132001}},
\href{http://www.arXiv.org/abs/1204.5201}{\texttt{arXiv:1204.5201}}.
%%CITATION = ARXIV:1204.5201;%%.

\bibitem{Czakon:2012zr}
\hrefCMSnoop {}{M.~Czakon and A.~Mitov, ``{NNLO corrections to top-pair
  production at hadron colliders: the all-fermionic scattering channels}'',}
  \textit{ JHEP} \textbf{ 12} (2012) 054,
  \href{http://dx.doi.org/10.1007/JHEP12(2012)054}{\doi{10.1007/JHEP12(2012)054}},
\href{http://www.arXiv.org/abs/1207.0236}{\texttt{arXiv:1207.0236}}.
%%CITATION = ARXIV:1207.0236;%%.

\bibitem{Czakon:2012pz}
\hrefCMSnoop {}{M.~Czakon and A.~Mitov, ``{NNLO corrections to top pair
  production at hadron colliders: the quark-gluon reaction}'',} \textit{ JHEP}
  \textbf{ 01} (2013) 080,
  \href{http://dx.doi.org/10.1007/JHEP01(2013)080}{\doi{10.1007/JHEP01(2013)080}},
\href{http://www.arXiv.org/abs/1210.6832}{\texttt{arXiv:1210.6832}}.
%%CITATION = ARXIV:1210.6832;%%.

\bibitem{Czakon:2013goa}
\hrefCMSnoop {}{M.~Czakon, P.~Fiedler, and A.~Mitov, ``{Total Top-Quark
  Pair-Production Cross Section at Hadron Colliders Through
  $O(\alpha_S^4)$}'',} \textit{ Phys. Rev. Lett.} \textbf{ 110} (2013) 252004,
  \href{http://dx.doi.org/10.1103/PhysRevLett.110.252004}{\doi{10.1103/PhysRevLett.110.252004}},
\href{http://www.arXiv.org/abs/1303.6254}{\texttt{arXiv:1303.6254}}.
%%CITATION = ARXIV:1303.6254;%%.

\bibitem{Gavin:2012sy}
\hrefCMSnoop {}{R.~Gavin, Y.~Li, F.~Petriello, and S.~Quackenbush, ``{W Physics
  at the LHC with FEWZ 2.1}'',} \textit{ Comput. Phys. Commun.} \textbf{ 184}
  (2013) 208,
  \href{http://dx.doi.org/10.1016/j.cpc.2012.09.005}{\doi{10.1016/j.cpc.2012.09.005}},
\href{http://www.arXiv.org/abs/1201.5896}{\texttt{arXiv:1201.5896}}.
%%CITATION = ARXIV:1201.5896;%%.

\bibitem{Gavin:2010az}
\hrefCMSnoop {}{R.~Gavin, Y.~Li, F.~Petriello, and S.~Quackenbush, ``{FEWZ 2.0:
  A code for hadronic Z production at next-to-next-to-leading order}'',}
  \textit{ Comput. Phys. Commun.} \textbf{ 182} (2011) 2388,
  \href{http://dx.doi.org/10.1016/j.cpc.2011.06.008}{\doi{10.1016/j.cpc.2011.06.008}},
\href{http://www.arXiv.org/abs/1011.3540}{\texttt{arXiv:1011.3540}}.
%%CITATION = ARXIV:1011.3540;%%.

\bibitem{Sjostrand:2014zea}
T.~Sj{\"o}strand\hrefCMSnoop {}{ {et~al.}, ``{An Introduction to PYTHIA
  8.2}'',} \textit{ Comput. Phys. Commun.} \textbf{ 191} (2015) 159,
  \href{http://dx.doi.org/10.1016/j.cpc.2015.01.024}{\doi{10.1016/j.cpc.2015.01.024}},
\href{http://www.arXiv.org/abs/1410.3012}{\texttt{arXiv:1410.3012}}.
%%CITATION = ARXIV:1410.3012;%%.

\bibitem{bib-nlo-nll-01}
\hrefCMSnoop {}{W.~Beenakker, R.~H{\"o}pker, M.~Spira, and P.~M. Zerwas,
  ``Squark and gluino production at hadron colliders'',} \textit{ Nucl. Phys.
  B} \textbf{ 492} (1997) 51,
  \href{http://dx.doi.org/10.1016/S0550-3213(97)00084-9}{\doi{10.1016/S0550-3213(97)00084-9}},
\href{http://www.arXiv.org/abs/hep-ph/9610490}{\texttt{arXiv:hep-ph/9610490}}.
%%CITATION = HEP-PH/9610490;%%.

\bibitem{bib-nlo-nll-02}
\hrefCMSnoop {}{A.~Kulesza and L.~Motyka, ``{Threshold resummation for
  squark-antisquark and gluino-pair production at the {LHC}}'',} \textit{ Phys.
  Rev. Lett.} \textbf{ 102} (2009) 111802,
  \href{http://dx.doi.org/10.1103/PhysRevLett.102.111802}{\doi{10.1103/PhysRevLett.102.111802}},
\href{http://www.arXiv.org/abs/0807.2405}{\texttt{arXiv:0807.2405}}.
%%CITATION = ARXIV:0807.2405;%%.

\bibitem{bib-nlo-nll-03}
\hrefCMSnoop {}{A.~Kulesza and L.~Motyka, ``{Soft gluon resummation for the
  production of gluino-gluino and squark-antisquark pairs at the {LHC}}'',}
  \textit{ Phys. Rev. D} \textbf{ 80} (2009) 095004,
  \href{http://dx.doi.org/10.1103/PhysRevD.80.095004}{\doi{10.1103/PhysRevD.80.095004}},
\href{http://www.arXiv.org/abs/0905.4749}{\texttt{arXiv:0905.4749}}.
%%CITATION = ARXIV:0905.4749;%%.

\bibitem{bib-nlo-nll-04}
W.~Beenakker\hrefCMSnoop {}{ {et~al.}, ``{Soft-gluon resummation for squark and
  gluino hadroproduction}'',} \textit{ JHEP} \textbf{ 12} (2009) 041,
  \href{http://dx.doi.org/10.1088/1126-6708/2009/12/041}{\doi{10.1088/1126-6708/2009/12/041}},
\href{http://www.arXiv.org/abs/0909.4418}{\texttt{arXiv:0909.4418}}.
%%CITATION = ARXIV:0909.4418;%%.

\bibitem{bib-nlo-nll-05}
W.~Beenakker\hrefCMSnoop {}{ {et~al.}, ``{Squark and gluino
  hadroproduction}'',} \textit{ Int. J. Mod. Phys. A} \textbf{ 26} (2011) 2637,
  \href{http://dx.doi.org/10.1142/S0217751X11053560}{\doi{10.1142/S0217751X11053560}},
\href{http://www.arXiv.org/abs/1105.1110}{\texttt{arXiv:1105.1110}}.
%%CITATION = ARXIV:1105.1110;%%.

\bibitem{Orbaker:2010zz}
\hrefCMSnoop {}{{CMS Collaboration}, ``{Fast simulation of the {CMS}
  detector}'',} \textit{ J. Phys. Conf. Ser.} \textbf{ 219} (2010) 032053,
\href{http://dx.doi.org/10.1088/1742-6596/219/3/032053}{\doi{10.1088/1742-6596/219/3/032053}}.
%%CITATION = 00462,219,032053;%%.

\bibitem{bib-cms-fastsim-02}
\href {http://cdsweb.cern.ch/record/1309890}{{CMS Collaboration}, ``Comparison
  of the fast simulation of {CMS} with the first {LHC} data'',} CMS Detector
  Performance Summary CMS-DP-2010-039, CERN, 2010.

\bibitem{Ball:2014uwa}
\hrefCMSnoop {}{{NNPDF} Collaboration, ``{Parton distributions for the LHC Run
  II}'',} \textit{ JHEP} \textbf{ 04} (2015) 040,
  \href{http://dx.doi.org/10.1007/JHEP04(2015)040}{\doi{10.1007/JHEP04(2015)040}},
\href{http://www.arXiv.org/abs/1410.8849}{\texttt{arXiv:1410.8849}}.
%%CITATION = ARXIV:1410.8849;%%.

\bibitem{Catani:2003zt}
\hrefCMSnoop {}{S.~Catani, D.~de~Florian, M.~Grazzini, and P.~Nason, ``{Soft
  gluon resummation for Higgs boson production at hadron colliders}'',}
  \textit{ JHEP} \textbf{ 07} (2003) 028,
  \href{http://dx.doi.org/10.1088/1126-6708/2003/07/028}{\doi{10.1088/1126-6708/2003/07/028}},
\href{http://www.arXiv.org/abs/hep-ph/0306211}{\texttt{arXiv:hep-ph/0306211}}.
%%CITATION = HEP-PH/0306211;%%.

\bibitem{Cacciari:2003fi}
M.~Cacciari\hrefCMSnoop {}{ {et~al.}, ``{The \ttbar cross-section at 1.8 TeV
  and 1.96 TeV: a study of the systematics due to parton densities and scale
  dependence}'',} \textit{ JHEP} \textbf{ 04} (2004) 068,
  \href{http://dx.doi.org/10.1088/1126-6708/2004/04/068}{\doi{10.1088/1126-6708/2004/04/068}},
\href{http://www.arXiv.org/abs/hep-ph/0303085}{\texttt{arXiv:hep-ph/0303085}}.
%%CITATION = HEP-PH/0303085;%%.

\bibitem{Chatrchyan:2013xna}
\hrefCMSnoop {}{{CMS Collaboration}, ``{Search for top-squark pair production
  in the single-lepton final state in pp collisions at $\sqrt{s}$ = 8 TeV}'',}
  \textit{ Eur. Phys. J. C} \textbf{ 73} (2013) 2677,
  \href{http://dx.doi.org/10.1140/epjc/s10052-013-2677-2}{\doi{10.1140/epjc/s10052-013-2677-2}},
\href{http://www.arXiv.org/abs/1308.1586}{\texttt{arXiv:1308.1586}}.
%%CITATION = ARXIV:1308.1586;%%.

\bibitem{Collaboration:2011ida}
\hrefCMSnoop {}{{CMS Collaboration}, ``{Search for new physics with jets and
  missing transverse momentum in pp collisions at $\sqrt{s}=7$ TeV}'',}
  \textit{ JHEP} \textbf{ 08} (2011) 155,
  \href{http://dx.doi.org/10.1007/JHEP08(2011)155}{\doi{10.1007/JHEP08(2011)155}},
\href{http://www.arXiv.org/abs/1106.4503}{\texttt{arXiv:1106.4503}}.
%%CITATION = ARXIV:1106.4503;%%.

\bibitem{Chatrchyan:2012lia}
\hrefCMSnoop {}{{CMS Collaboration}, ``{Search for new physics in the multijet
  and missing transverse momentum final state in proton-proton collisions at
  $\sqrt{s} = 7$ TeV}'',} \textit{ Phys. Rev. Lett.} \textbf{ 109} (2012)
  171803,
  \href{http://dx.doi.org/10.1103/PhysRevLett.109.171803}{\doi{10.1103/PhysRevLett.109.171803}},
\href{http://www.arXiv.org/abs/1207.1898}{\texttt{arXiv:1207.1898}}.
%%CITATION = ARXIV:1207.1898;%%.

\bibitem{PDG2014}
\hrefCMSnoop {}{{Particle Data Group}, K.~A. Olive {et~al.}, ``{Review of
  particle physics}'',} \textit{ Chin. Phys. C} \textbf{ 38} (2014) 090001,
\href{http://dx.doi.org/10.1088/1674-1137/38/9/090001}{\doi{10.1088/1674-1137/38/9/090001}}.
%%CITATION = PHRVA,D86,010001;%%.

\bibitem{cms_ttZ}
\hrefCMSnoop {}{{CMS Collaboration}, ``{ Observation of top quark pairs
  produced in association with a vector boson in pp collisions at $\sqrt{s}= 8$
  TeV}'',} \textit{ JHEP} \textbf{ 01} (2016) 096,
  \href{http://dx.doi.org/10.1007/JHEP01(2016)096}{\doi{10.1007/JHEP01(2016)096}},
\href{http://www.arXiv.org/abs/1510.01131}{\texttt{arXiv:1510.01131}}.
%%CITATION = ARXIV:1510.01131;%%.

\bibitem{Chatrchyan:2012rg}
\hrefCMSnoop {}{{CMS Collaboration}, ``{Search for supersymmetry in events with
  b-quark jets and missing transverse energy in pp collisions at 7 TeV}'',}
  \textit{ Phys. Rev. D} \textbf{ 86} (2012) 072010,
  \href{http://dx.doi.org/10.1103/PhysRevD.86.072010}{\doi{10.1103/PhysRevD.86.072010}},
\href{http://www.arXiv.org/abs/1208.4859}{\texttt{arXiv:1208.4859}}.
%%CITATION = ARXIV:1208.4859;%%.

\bibitem{Cowan:2010js}
\hrefCMSnoop {}{G.~Cowan, K.~Cranmer, E.~Gross, and O.~Vitells, ``Asymptotic
  formulae for likelihood-based tests of new physics'',} \textit{ Eur. Phys. J.
  C} \textbf{ 71} (2011) 1554,
  \href{http://dx.doi.org/10.1140/epjc/s10052-011-1554-0}{\doi{10.1140/epjc/s10052-011-1554-0}},
  \href{http://www.arXiv.org/abs/1007.1727}{\texttt{arXiv:1007.1727}}.
[Erratum: \DOI{10.1140/epjc/s10052-013-2501-z}].
%%CITATION = ARXIV:1007.1727;%%.

\bibitem{Junk1999}
\hrefCMSnoop {}{T.~Junk, ``{Confidence level computation for combining searches
  with small statistics}'',} \textit{ Nucl. Instr. and Meth. A} \textbf{ 434}
  (1999) 435,
  \href{http://dx.doi.org/10.1016/S0168-9002(99)00498-2}{\doi{10.1016/S0168-9002(99)00498-2}},
\href{http://www.arXiv.org/abs/hep-ex/9902006}{\texttt{arXiv:hep-ex/9902006}}.
%%CITATION = HEP-EX/9902006;%%.

\bibitem{bib-cls}
\hrefCMSnoop {}{A.~L. Read, ``Presentation of search results: the ${CL}_s$
  technique'',} \textit{ J. Phys. G} \textbf{ 28} (2002) 2693,
\href{http://dx.doi.org/10.1088/0954-3899/28/10/313}{\doi{10.1088/0954-3899/28/10/313}}.
%%CITATION = INSPIRE-599622;%%.

\bibitem{cms-note-2011-005}
\href {http://cdsweb.cern.ch/record/1379837}{{ATLAS and CMS Collaborations},
  ``{Procedure for the LHC Higgs boson search combination in Summer 2011}'',}
  Technical Report CMS-NOTE-2011-005, ATL-PHYS-PUB-2011-11, 2011.

\bibitem{Borschensky:2014cia}
C.~Borschensky\hrefCMSnoop {}{ {et~al.}, ``{Squark and gluino production cross
  sections in pp collisions at $\sqrt{s}$ = 13, 14, 33 and 100 TeV}'',}
  \textit{ Eur. Phys. J. C} \textbf{ 74} (2014) 3174,
  \href{http://dx.doi.org/10.1140/epjc/s10052-014-3174-y}{\doi{10.1140/epjc/s10052-014-3174-y}},
\href{http://www.arXiv.org/abs/1407.5066}{\texttt{arXiv:1407.5066}}.
%%CITATION = ARXIV:1407.5066;%%.

\end{thebibliography}\endgroup
\clearpage
\numberwithin{table}{section}
\appendix
\section{Selection efficiency for example signal models}
\label{sec:sel-eff}

\newcolumntype{R}{>{$}r<{$}}
\newcolumntype{L}{>{$}l<{$}}
\newcolumntype{M}{L@{$\;$}L}
\newcolumntype{S}{r@{$\,\pm\,$}r}
\begin{table*}[htb]
\topcaption{Absolute cumulative efficiencies in \% for each step of the event selection process,
listed for three representative signal models and choices for the gluino and LSP masses.
Only statistical uncertainties are shown.}
\centering
    \begin{tabular}{MSSS}
\hline
\multicolumn{2}{l}{Selection} & \multicolumn{2}{r}{$\Pp\Pp \to \PSg\PSg, \PSg \to \bbbar \PSGczDo$} & \multicolumn{2}{r}{$\Pp\Pp \to \PSg\PSg, \PSg \to \ttbar \PSGczDo$} & \multicolumn{2}{r}{$\Pp\Pp \to \PSg\PSg, \PSg \to \qqbar \PSGczDo$} \\
\multicolumn{2}{c}{}          & \multicolumn{2}{r}{$m_{\PSg}=1500\GeV$}                                       & \multicolumn{2}{r}{$m_{\PSg}=1500\GeV$}                                       & \multicolumn{2}{r}{$m_{\PSg}=1000\GeV$} \\
\multicolumn{2}{c}{}          & \multicolumn{2}{r}{$m_{\PSGczDo}=100\GeV$}                                    & \multicolumn{2}{r}{$m_{\PSGczDo}=100\GeV$}                                    & \multicolumn{2}{r}{$m_{\PSGczDo}=800\GeV$} \\
\hline
\njets                                & \geq4            & \hphantom{3333333}96.49 & 0.08 & \hphantom{333333}99.96 & 0.01 & \hphantom{3333333}76.87 & 0.14 \\
\HT                                   & >500\GeV         & 96.46 & 0.08 & 99.89 & 0.01 & 38.30 & 0.16 \\
\MHT                                  & >200\GeV         & 87.21 & 0.15 & 88.65 & 0.10 & 24.46 & 0.14 \\
\nmuons                               & =0               & 86.59 & 0.15 & 56.00 & 0.15 & 24.42 & 0.14 \\
\neles                                & =0               & 85.95 & 0.15 & 35.21 & 0.15 & 24.26 & 0.14 \\
\nisomuons                            & =0               & 85.66 & 0.15 & 34.46 & 0.15 & 24.19 & 0.14 \\
\nisoeles                             & =0               & 85.17 & 0.16 & 33.50 & 0.15 & 24.00 & 0.14 \\
\nisohads                             & =0               & 84.20 & 0.16 & 31.57 & 0.14 & 23.26 & 0.14 \\
\dphimht                              & >0.5,0.5,0.3,0.3 & 62.00 & 0.21 & 23.96 & 0.13 & 17.66 & 0.12 \\
\hline
\end{tabular}
  \label{tab:sel-eff}
\end{table*}
\clearpage

\section{Prefit background predictions}
\label{sec:prefit}

\begin{table*}[htb]
\renewcommand{\arraystretch}{1.40} \centering
\label{tab:pre-fit-results-nj1}
\topcaption{
Observed numbers of events and prefit background predictions for $4\leq\njets\leq6$.
These results are displayed in the leftmost section of Fig.~\ref{fig:fit-results}.
The first uncertainty is statistical and the second systematic.
}
\resizebox{\textwidth}{!}{
\begin{tabular}{ cccc|cccc|cc }
\hline
Bin & \MHT [\GeVns{}] & \HT [\GeVns{}] & \nbjets & Lost-$\Pe/\PGm$ & $\tau\to\text{had}$ & $\Z\to\PGn\PAGn$ & QCD & Total Pred. & Obs. \\ \hline
1 & 200-500 & 500-800 & 0 & $319\pm12\pm29$ & $310\pm11\pm19$ & $630\pm13^{+100}_{-80}$ & $220\pm4\pm110$ & $1480\pm26^{+150}_{-140}$ & 1602 \\
2 & 200-500 & 800-1200 & 0 & $59.2\pm4.3\pm5.4$ & $69.1\pm5.2\pm5.7$ & $145\pm6^{+26}_{-20}$ & $100\pm2\pm34$ & $373\pm12\pm42$ & 390 \\
3 & 200-500 & 1200+ & 0 & $13.8\pm2.2\pm1.4$ & $14.4\pm2.5\pm1.6$ & $31\pm3^{+12}_{-8}$ & $90\pm2\pm24$ & $150\pm6\pm27$ & 149 \\
4 & 500-750 & 500-1200 & 0 & $11.5\pm1.8\pm1.6$ & $8.9\pm1.7\pm1.3$ & $62\pm4^{+18}_{-13}$ & $0.38^{+0.12+0.42}_{-0.09-0.29}$ & $82\pm6^{+19}_{-13}$ & 120 \\
5 & 500-750 & 1200+ & 0 & $2.0\pm1.0\pm0.5$ & $0.56^{+0.52}_{-0.25}\pm0.15$ & $5.5\pm1.3^{+2.1}_{-1.5}$ & $1.0\pm0.2\pm0.9$ & $8.9^{+2.1+2.4}_{-1.8-1.8}$ & 13 \\
6 & 750+ & 800+ & 0 & $1.39^{+0.93}_{-0.77}\pm0.24$ & $1.77^{+0.99}_{-0.88}\pm0.34$ & $10.4\pm1.8^{+5.8}_{-4.1}$ & $0.24^{+0.09+0.26}_{-0.06-0.18}$ & $13.8\pm2.6^{+5.8}_{-4.1}$ & 12 \\
7 & 200-500 & 500-800 & 1 & $171\pm8\pm17$ & $206\pm9\pm13$ & $127\pm21\pm29$ & $69\pm2\pm37$ & $574\pm27\pm52$ & 499 \\
8 & 200-500 & 800-1200 & 1 & $31.4\pm4.0\pm3.0$ & $30.4\pm3.2\pm2.0$ & $29.2\pm4.9^{+7.4}_{-6.7}$ & $36\pm1\pm14$ & $127\pm9\pm16$ & 123 \\
9 & 200-500 & 1200+ & 1 & $6.3\pm1.7\pm0.8$ & $8.9\pm2.0\pm0.9$ & $6.3\pm1.2^{+2.7}_{-2.0}$ & $32\pm1\pm11$ & $54\pm4\pm11$ & 44 \\
10 & 500-750 & 500-1200 & 1 & $3.1\pm1.1\pm0.6$ & $2.64^{+0.96}_{-0.85}\pm0.48$ & $12.4\pm2.2^{+4.3}_{-3.5}$ & $0.07^{+0.04+0.09}_{-0.02-0.05}$ & $18.2\pm3.0^{+4.4}_{-3.6}$ & 22 \\
11 & 500-750 & 1200+ & 1 & $0.00^{+0.52}_{-0.00}\pm0.00$ & $0.07^{+0.46}_{-0.04}\pm0.02$ & $1.10\pm0.32^{+0.47}_{-0.36}$ & $0.38^{+0.12+0.41}_{-0.09-0.29}$ & $1.6^{+1.0}_{-0.3}\pm0.5$ & 1 \\
12 & 750+ & 800+ & 1 & $0.00^{+0.50}_{-0.00}\pm0.00$ & $0.54^{+0.56}_{-0.32}\pm0.13$ & $2.1\pm0.5^{+1.2}_{-0.9}$ & $0.02^{+0.06+0.06}_{-0.00-0.02}$ & $2.6^{+1.2+1.2}_{-0.6-0.9}$ & 2 \\
13 & 200-500 & 500-800 & 2 & $71.9\pm6.1^{+7.2}_{-6.7}$ & $77.2\pm5.0\pm5.4$ & $28\pm8\pm12$ & $15.9\pm1.1\pm8.8$ & $193\pm14\pm17$ & 202 \\
14 & 200-500 & 800-1200 & 2 & $18.8\pm4.8^{+2.5}_{-2.2}$ & $17.3\pm2.7\pm1.3$ & $6.4\pm1.9\pm2.9$ & $9.5\pm0.6\pm3.8$ & $52.0\pm7.7\pm5.4$ & 45 \\
15 & 200-500 & 1200+ & 2 & $2.1\pm1.1\pm0.2$ & $3.3\pm1.3\pm0.3$ & $1.39\pm0.42\pm0.73$ & $5.6\pm0.5\pm2.0$ & $12.3\pm2.5\pm2.2$ & 15 \\
16 & 500-750 & 500-1200 & 2 & $1.9\pm1.7^{+0.7}_{-0.2}$ & $2.26\pm0.88\pm0.86$ & $2.7\pm0.8\pm1.4$ & $0.03^{+0.02+0.04}_{-0.01-0.02}$ & $6.9\pm2.8\pm1.6$ & 5 \\
17 & 500-750 & 1200+ & 2 & $3.3\pm3.4^{+1.4}_{-0.0}$ & $0.07^{+0.46+0.02}_{-0.05-0.01}$ & $0.24\pm0.09\pm0.13$ & $0.07^{+0.08+0.09}_{-0.04-0.03}$ & $3.7^{+3.8+0.2}_{-3.4-0.1}$ & 0 \\
18 & 750+ & 800+ & 2 & $0.00^{+0.46}_{-0.00}\pm0.00$ & $0.04^{+0.46+0.02}_{-0.03-0.01}$ & $0.46\pm0.15^{+0.32}_{-0.26}$ & $0.03^{+0.06+0.05}_{-0.02-0.01}$ & $0.53^{+0.93+0.32}_{-0.16-0.26}$ & 1 \\
19 & 200-500 & 500-800 & 3+ & $6.3\pm1.7\pm0.8$ & $10.8\pm2.2\pm1.6$ & $6.5\pm3.8\pm2.9$ & $1.21^{+0.37}_{-0.29}\pm0.82$ & $24.8\pm5.4\pm3.5$ & 17 \\
20 & 200-500 & 800-1200 & 3+ & $0.24^{+0.67+0.03}_{-0.24-0.00}$ & $1.10^{+0.61}_{-0.40}\pm0.15$ & $1.49\pm0.87^{+0.70}_{-0.62}$ & $0.70^{+0.20}_{-0.16}\pm0.37$ & $3.5^{+1.6}_{-1.1}\pm0.8$ & 7 \\
21 & 200-500 & 1200+ & 3+ & $0.80^{+0.91}_{-0.57}\pm0.13$ & $0.11^{+0.46}_{-0.05}\pm0.02$ & $0.32\pm0.19^{+0.19}_{-0.13}$ & $0.72^{+0.23}_{-0.18}\pm0.36$ & $2.0^{+1.4}_{-0.7}\pm0.4$ & 3 \\
22 & 500-750 & 500-1200 & 3+ & $0.00^{+0.63}_{-0.00}\pm0.00$ & $0.03^{+0.46}_{-0.01}\pm0.01$ & $0.63\pm0.37^{+0.33}_{-0.26}$ & $0.05^{+0.11+0.09}_{-0.04-0.01}$ & $0.7^{+1.2}_{-0.4}\pm0.3$ & 0 \\
23 & 500-750 & 1200+ & 3+ & $0.00^{+0.77}_{-0.00}\pm0.00$ & $0.00^{+0.46}_{-0.00}\pm0.00$ & $0.06\pm0.04^{+0.03}_{-0.02}$ & $0.00^{+0.05+0.02}_{-0.00-0.00}$ & $0.1^{+1.2+0.1}_{-0.1-0.0}$ & 0 \\
24 & 750+ & 800+ & 3+ & $0.00^{+0.58}_{-0.00}\pm0.00$ & $0.00^{+0.46}_{-0.00}\pm0.00$ & $0.11\pm0.06^{+0.08}_{-0.04}$ & $0.00^{+0.04+0.02}_{-0.00-0.00}$ & $0.1^{+1.0+0.1}_{-0.1-0.0}$ & 0 \\
\hline
\end{tabular}
}
\end{table*}

\begin{table*}[htb]
\renewcommand{\arraystretch}{1.40} \centering
\label{tab:pre-fit-results-nj2}
\caption{
Observed numbers of events and prefit background predictions for $7\leq\njets\leq8$.
These results are displayed in the central section of Fig.~\ref{fig:fit-results}.
The first uncertainty is statistical and the second systematic.
}
\resizebox{\textwidth}{!}{
\begin{tabular}{ cccc|cccc|cc }
\hline
Bin & \MHT [\GeVns{}] & \HT [\GeVns{}] & \nbjets & Lost-$\Pe/\PGm$ & $\tau\to\text{had}$ & $\Z\to\PGn\PAGn$ & QCD & Total Pred. & Obs. \\ \hline
25 & 200-500 & 500-800 & 0 & $18.8\pm3.1\pm2.3$ & $24.5\pm2.7\pm2.0$ & $27.4\pm2.8^{+6.7}_{-5.1}$ & $14.1\pm1.6\pm8.2$ & $85\pm7\pm11$ & 85 \\
26 & 200-500 & 800-1200 & 0 & $12.5\pm1.8\pm2.2$ & $15.6\pm2.3\pm1.3$ & $17.3\pm2.3^{+4.2}_{-3.2}$ & $16.3\pm1.2\pm7.1$ & $61.7\pm4.8\pm8.4$ & 60 \\
27 & 200-500 & 1200+ & 0 & $2.9\pm1.2\pm0.3$ & $3.5\pm1.3\pm0.3$ & $6.0\pm1.3^{+2.3}_{-1.7}$ & $23.0\pm1.6\pm8.8$ & $35.4\pm3.1\pm9.0$ & 42 \\
28 & 500-750 & 500-1200 & 0 & $0.53^{+0.45}_{-0.26}\pm0.13$ & $0.81^{+0.66}_{-0.47}\pm0.19$ & $0.36\pm0.36^{+0.12}_{-0.00}$ & $0.06^{+0.10+0.09}_{-0.04-0.02}$ & $1.8^{+1.2}_{-0.8}\pm0.3$ & 1 \\
29 & 500-750 & 1200+ & 0 & $1.03^{+0.88+0.33}_{-0.80-0.24}$ & $1.44^{+0.93}_{-0.80}\pm0.29$ & $0.60\pm0.43^{+0.26}_{-0.18}$ & $0.26^{+0.17+0.30}_{-0.11-0.15}$ & $3.3\pm1.8\pm0.5$ & 1 \\
30 & 750+ & 800+ & 0 & $0.17^{+0.38+0.09}_{-0.17-0.00}$ & $0.17^{+0.49+0.11}_{-0.17-0.00}$ & $0.56\pm0.40^{+0.34}_{-0.16}$ & $0.19^{+0.16+0.23}_{-0.09-0.10}$ & $1.09^{+0.97+0.41}_{-0.53-0.19}$ & 1 \\
31 & 200-500 & 500-800 & 1 & $25.8\pm2.9\pm3.1$ & $31.8\pm3.0\pm2.3$ & $11.7\pm2.2\pm3.7$ & $8.1\pm1.3\pm5.1$ & $77.3\pm6.4\pm7.3$ & 63 \\
32 & 200-500 & 800-1200 & 1 & $9.0\pm1.6\pm1.2$ & $14.4\pm2.0\pm1.4$ & $7.4\pm1.5\pm2.3$ & $7.6\pm0.8\pm3.7$ & $38.3\pm4.0\pm4.7$ & 43 \\
33 & 200-500 & 1200+ & 1 & $3.3\pm1.1\pm0.4$ & $6.3\pm1.5\pm0.7$ & $2.6\pm0.7\pm1.1$ & $13.7\pm1.2\pm5.9$ & $25.9\pm2.9\pm6.1$ & 29 \\
34 & 500-750 & 500-1200 & 1 & $0.46^{+0.49}_{-0.27}\pm0.11$ & $0.51^{+0.55}_{-0.29}\pm0.11$ & $0.15\pm0.16^{+0.06}_{-0.00}$ & $0.00^{+0.12+0.05}_{-0.00-0.00}$ & $1.1^{+1.1}_{-0.6}\pm0.2$ & 2 \\
35 & 500-750 & 1200+ & 1 & $0.00^{+0.40}_{-0.00}\pm0.00$ & $0.25^{+0.49}_{-0.18}\pm0.05$ & $0.26\pm0.19^{+0.12}_{-0.07}$ & $0.12^{+0.14+0.16}_{-0.07-0.05}$ & $0.63^{+0.92+0.21}_{-0.27-0.10}$ & 2 \\
36 & 750+ & 800+ & 1 & $0.00^{+0.45}_{-0.00}\pm0.00$ & $0.02^{+0.46+0.01}_{-0.01-0.00}$ & $0.24\pm0.17^{+0.15}_{-0.07}$ & $0.00^{+0.08+0.03}_{-0.00-0.00}$ & $0.25^{+0.93+0.16}_{-0.17-0.07}$ & 1 \\
37 & 200-500 & 500-800 & 2 & $13.2\pm2.2\pm1.5$ & $16.0\pm1.9\pm1.2$ & $4.8\pm1.5\pm2.4$ & $0.16^{+0.32+0.57}_{-0.00-0.16}$ & $34.1\pm4.3\pm3.1$ & 32 \\
38 & 200-500 & 800-1200 & 2 & $6.3\pm1.3\pm0.7$ & $10.7\pm1.8\pm0.9$ & $3.0\pm1.0\pm1.5$ & $2.2\pm.5\pm1.1$ & $22.2\pm3.3\pm2.2$ & 17 \\
39 & 200-500 & 1200+ & 2 & $1.73^{+0.79}_{-0.62}\pm0.20$ & $1.89^{+0.88}_{-0.75}\pm0.18$ & $1.06\pm0.38\pm0.60$ & $3.6\pm0.6\pm1.6$ & $8.2\pm1.7\pm1.8$ & 4 \\
40 & 500-750 & 500-1200 & 2 & $0.00^{+0.39}_{-0.00}\pm0.00$ & $0.04^{+0.46}_{-0.02}\pm0.01$ & $0.06^{+0.07+0.03}_{-0.06-0.00}$ & $0.00^{+0.12+0.05}_{-0.00-0.00}$ & $0.10^{+0.86+0.06}_{-0.06-0.01}$ & 0 \\
41 & 500-750 & 1200+ & 2 & $0.00^{+0.43}_{-0.00}\pm0.00$ & $0.07^{+0.47+0.04}_{-0.07-0.00}$ & $0.11\pm0.08^{+0.06}_{-0.02}$ & $0.03^{+0.11+0.05}_{-0.02-0.01}$ & $0.21^{+0.90+0.08}_{-0.11-0.03}$ & 1 \\
42 & 750+ & 800+ & 2 & $0.00^{+0.34}_{-0.00}\pm0.00$ & $0.13^{+0.48+0.06}_{-0.13-0.00}$ & $0.10\pm0.07^{+0.07}_{-0.02}$ & $0.00^{+0.08+0.03}_{-0.00-0.00}$ & $0.23^{+0.82+0.08}_{-0.15-0.02}$ & 0 \\
43 & 200-500 & 500-800 & 3+ & $3.9\pm1.2\pm0.5$ & $5.8\pm1.3\pm0.7$ & $2.5\pm1.5^{+1.8}_{-1.0}$ & $1.09^{+0.62+0.86}_{-0.41-0.68}$ & $13.3\pm3.0^{+2.1}_{-1.5}$ & 3 \\
44 & 200-500 & 800-1200 & 3+ & $0.44^{+0.49}_{-0.25}\pm0.05$ & $1.66^{+0.76}_{-0.60}\pm0.26$ & $1.6\pm1.0^{+1.1}_{-0.7}$ & $0.60^{+0.30}_{-0.21}\pm0.39$ & $4.3^{+1.6+1.2}_{-1.3-0.8}$ & 4 \\
45 & 200-500 & 1200+ & 3+ & $0.66^{+0.72}_{-0.52}\pm0.12$ & $0.65^{+0.61}_{-0.40}\pm0.10$ & $0.56\pm0.35^{+0.42}_{-0.21}$ & $0.04^{+0.19+0.12}_{-0.00-0.04}$ & $1.9^{+1.4+0.5}_{-1.0-0.3}$ & 1 \\
46 & 500-750 & 500-1200 & 3+ & $0.00^{+0.52}_{-0.00}\pm0.00$ & $0.00^{+0.46}_{-0.00}\pm0.00$ & $0.03^{+0.04+0.02}_{-0.03-0.00}$ & $0.04^{+0.09+0.07}_{-0.03-0.01}$ & $0.07^{+0.98+0.07}_{-0.05-0.01}$ & 0 \\
47 & 500-750 & 1200+ & 3+ & $0.00^{+0.47}_{-0.00}\pm0.00$ & $0.00^{+0.46}_{-0.00}\pm0.00$ & $0.06\pm0.05^{+0.04}_{-0.00}$ & $0.00^{+0.09+0.03}_{-0.00-0.00}$ & $0.06^{+0.94+0.05}_{-0.05-0.00}$ & 0 \\
48 & 750+ & 800+ & 3+ & $0.00^{+0.61}_{-0.00}\pm0.00$ & $0.01^{+0.46+0.01}_{-0.01-0.00}$ & $0.05\pm0.05^{+0.05}_{-0.00}$ & $0.00^{+0.08+0.03}_{-0.00-0.00}$ & $0.1^{+1.1+0.1}_{-0.1-0.0}$ & 0 \\
\hline
\end{tabular}
}
\end{table*}

\begin{table*}[htb]
\renewcommand{\arraystretch}{1.40} \centering
\label{tab:pre-fit-results-nj3}
\caption{
Observed numbers of events and prefit background predictions for $\njets\geq9$.
These results are displayed in the rightmost section of Fig.~\ref{fig:fit-results}.
The first uncertainty is statistical and the second systematic.
}
\resizebox{\textwidth}{!}{
\begin{tabular}{ cccc|cccc|cc }
\hline
Bin & \MHT [\GeVns{}] & \HT [\GeVns{}] & \nbjets & Lost-$\Pe/\PGm$ & $\tau\to\text{had}$ & $\Z\to\PGn\PAGn$ & QCD & Total Pred. & Obs. \\ \hline
49 & 200-500 & 500-800 & 0 & $0.99^{+0.59}_{-0.45}\pm0.21$ & $0.61^{+0.52}_{-0.23}\pm0.09$ & $0.26\pm0.26^{+0.12}_{-0.00}$ & $0.92^{+0.54+0.80}_{-0.35-0.57}$ & $2.8^{+1.3}_{-0.8}\pm0.7$ & 2 \\
50 & 200-500 & 800-1200 & 0 & $2.12^{+0.72}_{-0.62}\pm0.33$ & $3.9\pm1.2\pm0.4$ & $2.14\pm0.81^{+0.81}_{-0.64}$ & $0.78^{+0.31}_{-0.23}\pm0.55$ & $9.0\pm2.0\pm1.1$ & 12 \\
51 & 200-500 & 1200+ & 0 & $0.58^{+0.54}_{-0.35}\pm0.08$ & $1.05^{+0.76}_{-0.61}\pm0.15$ & $0.42\pm0.30^{+0.18}_{-0.12}$ & $3.9\pm0.7\pm2.5$ & $6.0^{+1.5}_{-1.2}\pm2.5$ & 8 \\
52 & 500-750 & 500-1200 & 0 & $0.00^{+0.34}_{-0.00}\pm0.00$ & $0.00^{+0.46}_{-0.00}\pm0.00$ & $0.15\pm0.15^{+0.11}_{-0.00}$ & $0.00^{+0.11+0.04}_{-0.00-0.00}$ & $0.15^{+0.82+0.11}_{-0.15-0.00}$ & 0 \\
53 & 500-750 & 1200+ & 0 & $0.14^{+0.36+0.05}_{-0.14-0.00}$ & $0.02^{+0.46+0.01}_{-0.02-0.00}$ & $0.00^{+0.76}_{-0.00}\pm0.00$ & $0.00^{+0.09+0.04}_{-0.00-0.00}$ & $0.2^{+1.1+0.1}_{-0.2-0.0}$ & 0 \\
54 & 750+ & 800+ & 0 & $0.00^{+0.28}_{-0.00}\pm0.00$ & $0.00^{+0.46}_{-0.00}\pm0.00$ & $0.00^{+0.79}_{-0.00}\pm0.00$ & $0.00^{+0.08+0.03}_{-0.00-0.00}$ & $0.0^{+1.1+0.1}_{-0.0-0.0}$ & 0 \\
55 & 200-500 & 500-800 & 1 & $1.36^{+0.66}_{-0.53}\pm0.19$ & $1.58^{+0.71}_{-0.54}\pm0.19$ & $0.19\pm0.19^{+0.10}_{-0.00}$ & $0.09^{+0.22+0.15}_{-0.07-0.02}$ & $3.2^{+1.4}_{-1.1}\pm0.3$ & 6 \\
56 & 200-500 & 800-1200 & 1 & $3.19^{+0.99}_{-0.91}\pm0.52$ & $4.1\pm1.2\pm0.4$ & $1.57\pm0.64\pm0.68$ & $0.88^{+0.34}_{-0.25}\pm0.64$ & $9.7\pm2.2\pm1.2$ & 4 \\
57 & 200-500 & 1200+ & 1 & $1.70^{+0.85}_{-0.73}\pm0.25$ & $1.41^{+0.79}_{-0.65}\pm0.25$ & $0.31\pm0.22^{+0.15}_{-0.08}$ & $2.4\pm0.5\pm1.6$ & $5.8\pm1.6\pm1.7$ & 3 \\
58 & 500-750 & 500-1200 & 1 & $0.00^{+0.40}_{-0.00}\pm0.00$ & $0.05^{+0.46+0.02}_{-0.05-0.00}$ & $0.11\pm0.11^{+0.08}_{-0.00}$ & $0.00^{+0.11+0.04}_{-0.00-0.00}$ & $0.16^{+0.88+0.09}_{-0.12-0.00}$ & 0 \\
59 & 500-750 & 1200+ & 1 & $0.00^{+0.41}_{-0.00}\pm0.00$ & $0.15^{+0.48+0.04}_{-0.14-0.00}$ & $0.00^{+0.66}_{-0.00}\pm0.00$ & $0.00^{+0.09+0.03}_{-0.00-0.00}$ & $0.2^{+1.1+0.1}_{-0.1-0.0}$ & 1 \\
60 & 750+ & 800+ & 1 & $0.00^{+0.33}_{-0.00}\pm0.00$ & $0.00^{+0.46}_{-0.00}\pm0.00$ & $0.00^{+0.68}_{-0.00}\pm0.00$ & $0.00^{+0.08+0.03}_{-0.00-0.00}$ & $0.0^{+1.1+0.1}_{-0.0-0.0}$ & 0 \\
61 & 200-500 & 500-800 & 2 & $1.38^{+0.74}_{-0.62}\pm0.18$ & $1.51^{+0.77}_{-0.61}\pm0.15$ & $0.10\pm0.10^{+0.07}_{-0.00}$ & $0.00^{+0.22+0.11}_{-0.00-0.00}$ & $3.0^{+1.5}_{-1.2}\pm0.3$ & 3 \\
62 & 200-500 & 800-1200 & 2 & $1.39^{+0.68}_{-0.57}\pm0.20$ & $2.20^{+0.92}_{-0.80}\pm0.20$ & $0.87\pm0.41^{+0.54}_{-0.46}$ & $0.26^{+0.22+0.24}_{-0.13-0.13}$ & $4.7^{+1.7}_{-1.4}\pm0.6$ & 1 \\
63 & 200-500 & 1200+ & 2 & $0.28^{+0.48}_{-0.20}\pm0.04$ & $1.40^{+0.83}_{-0.70}\pm0.19$ & $0.17\pm0.13^{+0.11}_{-0.04}$ & $1.38^{+0.45}_{-0.35}\pm0.95$ & $3.2^{+1.4}_{-1.0}\pm1.0$ & 2 \\
64 & 500-750 & 500-1200 & 2 & $0.00^{+0.36}_{-0.00}\pm0.00$ & $0.00^{+0.46}_{-0.00}\pm0.00$ & $0.06\pm0.06^{+0.05}_{-0.00}$ & $0.00^{+0.11+0.04}_{-0.00-0.00}$ & $0.06^{+0.83+0.07}_{-0.06-0.00}$ & 0 \\
65 & 500-750 & 1200+ & 2 & $0.00^{+0.45}_{-0.00}\pm0.00$ & $0.01^{+0.46}_{-0.01}\pm0.00$ & $0.00^{+0.52}_{-0.00}\pm0.00$ & $0.00^{+0.09+0.03}_{-0.00-0.00}$ & $0.0^{+1.1+0.1}_{-0.0-0.0}$ & 0 \\
66 & 750+ & 800+ & 2 & $0.00^{+0.43}_{-0.00}\pm0.00$ & $0.00^{+0.46}_{-0.00}\pm0.00$ & $0.00^{+0.52}_{-0.00}\pm0.00$ & $0.00^{+0.08+0.03}_{-0.00-0.00}$ & $0.0^{+1.0+0.1}_{-0.0-0.0}$ & 0 \\
67 & 200-500 & 500-800 & 3+ & $0.30^{+0.48}_{-0.21}\pm0.05$ & $1.13^{+0.79}_{-0.64}\pm0.16$ & $0.02^{+0.03+0.03}_{-0.02-0.00}$ & $0.00^{+0.22+0.09}_{-0.00-0.00}$ & $1.5^{+1.3}_{-0.9}\pm0.2$ & 0 \\
68 & 200-500 & 800-1200 & 3+ & $1.9\pm1.4\pm0.3$ & $0.70^{+0.60}_{-0.38}\pm0.09$ & $0.18\pm0.13^{+0.24}_{-0.06}$ & $0.27^{+0.22+0.25}_{-0.13-0.14}$ & $3.1^{+2.0}_{-1.7}\pm0.5$ & 1 \\
69 & 200-500 & 1200+ & 3+ & $0.46^{+0.64+0.06}_{-0.46-0.00}$ & $0.32^{+0.54}_{-0.28}\pm0.04$ & $0.04\pm0.03^{+0.05}_{-0.00}$ & $0.04^{+0.10+0.07}_{-0.03-0.01}$ & $0.9^{+1.2+0.1}_{-0.8-0.0}$ & 0 \\
70 & 500-750 & 500-1200 & 3+ & $0.13^{+0.47+0.05}_{-0.13-0.00}$ & $0.00^{+0.46}_{-0.00}\pm0.00$ & $0.01^{+0.02+0.02}_{-0.01-0.00}$ & $0.00^{+0.11+0.04}_{-0.00-0.00}$ & $0.14^{+0.93+0.04}_{-0.13-0.00}$ & 0 \\
71 & 500-750 & 1200+ & 3+ & $0.00^{+0.41}_{-0.00}\pm0.00$ & $0.00^{+0.46}_{-0.00}\pm0.00$ & $0.00^{+0.30}_{-0.00}\pm0.00$ & $0.00^{+0.09+0.02}_{-0.00-0.00}$ & $0.00^{+0.93+0.02}_{-0.00-0.00}$ & 0 \\
72 & 750+ & 800+ & 3+ & $0.00^{+0.44}_{-0.00}\pm0.00$ & $0.00^{+0.46}_{-0.00}\pm0.00$ & $0.00^{+0.28}_{-0.00}\pm0.00$ & $0.00^{+0.08+0.03}_{-0.00-0.00}$ & $0.00^{+0.95+0.03}_{-0.00-0.00}$ & 0 \\
\hline
\end{tabular}
}
\end{table*}
\cleardoublepage \section{The CMS Collaboration \label{app:collab}}\begin{sloppypar}\hyphenpenalty=5000\widowpenalty=500\clubpenalty=5000\textbf{Yerevan Physics Institute,  Yerevan,  Armenia}\\*[0pt]
V.~Khachatryan, A.M.~Sirunyan, A.~Tumasyan
\vskip\cmsinstskip
\textbf{Institut f\"{u}r Hochenergiephysik der OeAW,  Wien,  Austria}\\*[0pt]
W.~Adam, E.~Asilar, T.~Bergauer, J.~Brandstetter, E.~Brondolin, M.~Dragicevic, J.~Er\"{o}, M.~Flechl, M.~Friedl, R.~Fr\"{u}hwirth\cmsAuthorMark{1}, V.M.~Ghete, C.~Hartl, N.~H\"{o}rmann, J.~Hrubec, M.~Jeitler\cmsAuthorMark{1}, A.~K\"{o}nig, I.~Kr\"{a}tschmer, D.~Liko, T.~Matsushita, I.~Mikulec, D.~Rabady, N.~Rad, B.~Rahbaran, H.~Rohringer, J.~Schieck\cmsAuthorMark{1}, J.~Strauss, W.~Treberer-Treberspurg, W.~Waltenberger, C.-E.~Wulz\cmsAuthorMark{1}
% ...
\vskip\cmsinstskip
% ...
1:~~Also at Vienna University of Technology, Vienna, Austria\\
% ...

\end{sloppypar}
\end{document}